%% file: main_EMSE.tex
\documentclass[smallextended]{svjour3} 
\smartqed
\usepackage{wrapfig}
\usepackage{graphicx}
\usepackage{multicol}
\usepackage{multirow}
\usepackage{hhline}
\usepackage{comment}
\usepackage{hyphenat}
\usepackage{subcaption}
\usepackage{float}
\usepackage{caption}
\usepackage{fancybox}
\usepackage{array}
\usepackage[toc,page]{appendix}
\usepackage{makecell}
\usepackage{rotating}
\usepackage{bigstrut}
\usepackage{booktabs}
\usepackage{subcaption}
\usepackage[misc]{ifsym}

\usepackage{changepage}
\usepackage{hyperref} 
\usepackage{soul} 
\usepackage[super]{nth}
\usepackage{adjustbox}
\usepackage{amsmath,amssymb,amsfonts}
\usepackage{algorithmic}
\usepackage{graphicx}
\usepackage{textcomp}
\usepackage{url}
\usepackage{relsize} 
\usepackage{amssymb} 

\newcolumntype{P}[1]{>{\centering\arraybackslash}p{#1}}

\usepackage[graphicx]{realboxes}
\usepackage[most]{tcolorbox}
\definecolor{custom-gray}{cmyk}{0, 0, 0, 0.7, 1.00}
\newtcbtheorem[no counter]{Summary}{\hskip-0.97em}{enhanced,drop shadow={black!50!white},
 coltitle=white,
 top=0.15in,
 attach boxed title to top left=
 {xshift=1.5em,yshift=-\tcboxedtitleheight/2},
 boxed title style={size=small,colback=custom-gray}
}{summary}

\definecolor{codegreen}{rgb}{0,0.6,0}
\definecolor{codegray}{rgb}{0.5,0.5,0.5}
\definecolor{codepurple}{rgb}{0.58,0,0.82}
\definecolor{backcolour}{rgb}{0.95,0.95,0.92}

\lstdefinestyle{mystyle}{
    backgroundcolor=\color{backcolour},   
    commentstyle=\color{codegreen},
    keywordstyle=\color{magenta},
    numberstyle=\tiny\color{codegray},
    stringstyle=\color{codepurple},
    basicstyle=\ttfamily\footnotesize,
    breakatwhitespace=false,         
    breaklines=true,                 
    captionpos=b,                    
    keepspaces=true,                 
    numbers=left,                    
    numbersep=5pt,                  
    showspaces=false,                
    showstringspaces=false,
    showtabs=false,                  
    tabsize=2
}

\newcommand\rqone[0]{RQ1: What is the prevalence of data quality antipatterns in software defect prediction data?}
    \newcommand\rqtwoone[0]{RQ2: Does the order of cleaning data quality antipatterns impact the model performance?}
\newcommand\rqtwotwo[0]{RQ3: How does data quality anti-pattern removal impact model performance?}
\newcommand{\rqthree}[0]{RQ4: Do models built from data containing different antipatterns give inconsistent interpretation results?}

\newcommand{\aadi}[1]{\textcolor{blue}{\textit{Aadi: [#1]}}}

\newcommand{\dayi}[1]{\textcolor{red}{\textit{Dayi: [#1]}}}

\begin{document}
\title{Data Quality Antipatterns for Software Analytics}
\subtitle{A Case Study of Software Defect Prediction}
 
\author{Aaditya~Bhatia \and
Dayi~Lin \and Gopi~Krishnan~Rajbahadur \and
Bram~Adams \and Ahmed~E.~Hassan}

\institute{Aaditya~Bhatia \at
Queen's University, Kingston, ON, Canada\\
\email{aaditya.bhatia@queensu.ca} \and 
Dayi~Lin \and Gopi~Krishnan~Rajbahadur \at {Centre for Software Excellence, Huawei Canada, Kingston, ON, Canada}\\
\email{\{dayi.lin, gopi.krishnan.rajbahadur1\}@huawei.com}
\and Bram Adams \and Ahmed E. Hassan \at
{Queen's University, Kingston, ON, Canada}\\
\email{ahmed, bram.adams@queensu.ca}\\
\and Aaditya Bhatia is the corresponding author\\
}
\date{Received: date / Accepted: date} 
\authorrunning{Bhatia et al.}
\titlerunning{Data Quality Antipatterns for ML}
\maketitle

\input{abstract}
\keywords {Data Quality \and Machine Learning}

\input{intro2}
\input{LiteratureReview}

\input{taxonomy}
\input{Appendix_mitigation_strategies}
\input{case_study}

\section{Results}\label{sec.results}
In this section, we delve into the motivation and results of each of our research questions.
    \input{rq1}\label{subsec.RQ1}
    \input{rq2_orders}\label{subsec.RQ2}
    \input{rq2_aps}\label{subsec.RQ3}
    \input{rq3}\label{subsec.RQ3}
\input{Threats}
\input{Discussion}
\input{Implications}
\input{Conclusion}


\bibliographystyle{IEEEtran}
\bibliography{references,zotero_references}
\end{document}

%% file: Abstract.tex
\begin{abstract}
\textit{Background}: Data quality is crucial in software analytics, particularly for machine learning (ML) applications like software defect prediction (SDP). Despite the widespread use of ML models in software engineering, the impact of data quality antipatterns on these models remains underexplored.

\textit{Objective}: This study aims to develop a comprehensive taxonomy of data quality antipatterns specific to ML and assess their impact on the performance and interpretation of software analytics models.

\textit{Methods}: Through a literature review, we identified eight types and 14 sub-types of ML-specific data quality antipatterns. We then conducted a series of experiments to determine the prevalence of data quality antipatterns in SDP data (RQ1), assess the impact of the order of cleaning antipatterns on model performance (RQ2), evaluate the effects of antipattern removal on model performance (RQ3), and examine the consistency of interpretation results from models built with different antipatterns (RQ4).

\textit{Results}: In our SDP case study, we identified the prevalence of nine antipatterns from our taxonomy. Notably, we observed that more than 90\% of the antipatterns identified from our taxonomy overlapped at both the row and column levels within the structured SDP dataset. This overlap can be a major issue as it can complicate the prioritization of cleaning processes as it may obscure critical antipatterns and lead to excessive data removal or improper cleaning. 
Next, the order of cleaning antipatterns significantly impacts the performance of ML models, with learners like neural networks being more resilient to changes in cleaning order than simpler models like logistic regression. Additionally, antipatterns such as Tailed Distributions and Class Overlap have a statistically significant correlation with performance metrics when other antipatterns are cleaned. Models built from data with different antipatterns showed moderate consistency in interpretation results.

\textit{Conclusion}: Our results indicate that the order of cleaning different antipatterns impact the performance of ML models. Five of the studied antipattens from our taxonomy have a statistically significant correlation with model performance in a setting where the remaining antipatterns have been cleaned out. 
Finally, apart from model performance, model interpretation results are also moderately impacted by different data quality antipatterns. 
\end{abstract}

%% file: intro2.tex
\section{Introduction}\label{sec:intro}

Machine Learning (ML) has been widely adopted in various domains of software analytics, including software defect prediction, technical debt identification, and effort estimation~\cite{menzies2010defect,tsoukalas2021machine,zimmermann2009cross,kamei2012large,ghotra2015revisiting}. These applications leverage ML to make key decisions, such as predicting future bugs, identifying problematic code, or estimating the required effort for software projects. In the domain of software analytics, data for models typically originates from software project repositories, issue tracking systems, and code commit histories, providing real-time and historical insights into software development practices. The users of these ML models expect not only high precision (for identification of specific issues) and recall (for ensuring no potential problems are overlooked), but also rely on these models for critical decision-making in software development and maintenance~\cite{tantithamthavorn2018experience,lyu2021empirical}. 

Apart from the choice of learning algorithm, hyperparameter tuning or embedding, the reliability of these decisions heavily depends on the quality of the data used to train the ML models~\cite{domingos2012few,tantithamthavorn2015impact,shepperd2013data}. Indeed, even if the code building the ML model is of high quality and is highly optimized, if the underlying data is flawed, the resulting model will be ineffective and potentially misleading. Unlike traditional software, where requirements are embedded as business logic encapsulated in software code, the requirements of ML models are embedded within the data itself. For example, to train a classifier distinguishing between cats and dogs, the training data must consist of accurate images of cats and dogs. If the dataset includes mislabeled images (e.g., an image of a rabbit labeled as a dog), the model will be compromised. Thus, the requirements of ML models are inherently tied to the quality and correctness of the data, making data an essential asset that determines the quality of the trained ML model~\cite{budach2022effects}.

Data being an important ML asset comes with its own suite of issues. While data quality issues are a recognized problem that compromises the quality of model training data, ``data quality antipatterns" are an important subset of such issues that could become problematic under certain circumstances. Similar to code antipatterns, their presence is not a direct indication of bugs or incorrect behavior, yet they can potentially compromise other qualities of ML models such as model performance and model interpretation. 
For example, class overlap occurs when samples with different classes have highly overlapping feature space, which may complicate their classification~\cite{kocaguneli2011exploiting}. Removing these points may help clarify class boundaries but could perhaps lose valuable information determining class boundaries, thereby compromising the model learning. Similarly, on the one hand, removing correlated \& redundant features reduces data (data loss can degrade performance~\cite{halevy2009unreasonable}), but on the other hand render interpretation results inaccurate~\cite{jiarpakdee2018autospearman}. Another example is the presence of outliers, extreme values deviating from other observations, can skew model training and generalization but might not impact resilient learners. 

Literature~\cite{shepperd2013data,sharma2022performance,breck2019data,vzliobaite2016overview,chen2018tackling,denil2010overlap,vuttipittayamongkol2021class,gong2019empirical,johnson2019survey,yang2020rethinking,ali2019imbalance,daoud2017multicollinearity} has identified that such antipatterns can significantly affect the performance of software analytics ML models. For instance, while checking the impact of different bug labeling techniques on the performance of defect prediction models, Yatish et al.~\cite{yatish2019mining} obtained a 3\% to 5\% boost in AUC by correcting labels for defect classification models. Similarly, Gong et al.~\cite{gong2019empirical} observed a slight boost in AUC by removing data points having overlapping classes.

Operationalizing data for training ML models involves various practical preprocessing decisions, including but not limited to ensuring the quality of the data used for training. While research has often observed the impact of data quality antipatterns in isolation, such scenarios with one isolated antipattern are rare in the real-world datasets. Particularly, the shortcomings of data quality research includes:

\begin{enumerate}
    \item \textbf{While multiple antipatterns exist in realistic datasets, calling the need for prioritization of critical antipatterns}, most research on data quality antipatterns neglects the fact that that multiple antipatterns can coexist and interact. For example, in the above example of defect classification by Yatish et al.~\cite{yatish2019mining}, the authors did not consider the impact of class overlap or class imbalance or tailed distributions, either separately or together. Given the existence a plethora of antipatterns~\cite{shepperd2013data,sharma2022performance,breck2019data,vzliobaite2016overview,chen2018tackling,denil2010overlap,vuttipittayamongkol2021class,gong2019empirical,johnson2019survey,yang2020rethinking,ali2019imbalance,daoud2017multicollinearity}, addressing each and every antipattern can not only be cumbersome, but also might not be required due to interactions and/or dependencies between antipatterns. Hence, researchers and practitioners need a better understanding of the critical antipatterns to address, to prevent burdensome or over-cleaning of their datasets.

    \item \textbf{Research has mostly focused on the ``harmful'' effects of some antipatterns, neglecting the reasons for retaining other antipatterns.} Not all antipatterns can be significantly harmful, but their removal will incur substantial costs- valuable data might be removed or imputed during the data cleaning processes. Given that some data is better than less data~\cite{halevy2009unreasonable}, determining if it is worth sacrificing data due to rigorous cleaning is vital especially in scenarios where data availability is sparse, the performance degradation costs can be non-trivial~\cite{chawla2002smote}. 

    An analogy can be drawn to the renowned research by Kapser et al.~\cite{kapser2008cloning}, which suggests that considering software code clones as “harmful” overlooks their benefits, and that deeming them outright “harmful” is, in itself, harmful. Essentially, software quality can be preserved by \textit{managing} code clones, rather than dismissing them outright.  Similarly, some data quality antipatterns may also require careful management rather than outright dismissal leading to over-cleaning.
    
    \item \textbf{The optimal order of antipattern cleaning is unknown.} When multiple antipatterns are addressed, the order in which cleanings are performed can significantly affect the quantity of data along with the content of the resulting dataset. This issue becomes more complex in distributed data pipelines, where data may be partially cleaned by devices like sensors before being transmitted to a central server for machine learning model development. In such cases, the data arrives at the server already partially processed, i.e., if a few antipattern cleanings are already done, it is crucial to correctly choose the subsequent antipatterns to be cleaned, since such choices may impact the performance of ML models. Practitioners must make informed decisions about the order of antipattern removal to ensure that the final dataset is properly addressed for optimal model performance.
    
    \item \textbf{The impact of antipattern removal on model interpretation is unknown.} Model interpretation is vital in software analytics as  it enables informed decision-making based on the model's predictions. In software defect prediction for instance, understanding which features contribute most to a prediction can help developers focus their debugging efforts more effectively. Similarly, in effort estimation, interpreting the model can reveal which aspects of a project are likely to require the most resources, enabling better project planning and resource allocation. Prior studies have explored the impact of various factors such as learner characteristics, project specifics, hyperparameter tuning, and sampling randomness on interpretation results~\cite{lyu2021towards}, while overlooking the inconsistency introduced by various data quality antipatterns on the interpretation results of ML models.

\end{enumerate}


Hence, to fill the gap in literature and help practitioners in making informed decisions while preprocessing data for ML building ML models, in this study, we use Software Defect Prediction (SDP) as a case study to understand the joint impact of these antipatterns on the performance and interpretation of software analytics models. This allows us to investigate these antipatterns in a practical, real-world context, and have a focused understanding of their implications on software analytics using a dataset widely used in research~\cite{rajapaksha2021sqaplanner,jiarpakdee2021practitioners,tong2019kernel,gong2021revisiting,moussa2022use,jiarpakdee2020empirical,wattanakriengkrai2020predicting,fu2022linevul}. SDP is particularly significant due to its widespread use in research and the criticality of its role in maintaining software quality. SDP datasets used in these studies have been reused extensively in the software analytics research community~\cite{menzies2010defect,jiarpakdee2021practitioners}, along with for ML benchmarking~\cite{ghotra2015revisiting,aleem2015benchmarking}. Therefore, understanding the impact of data quality antipatterns in the context of SDP can provide valuable insights that can be generalized to other areas of software analytics. 

Using SDP as our use-case, we determine whether the order of cleaning antipatterns could potentially affect the performance of the models, and, if so, find the optimal sequence for a more efficient and effective cleaning process. Second, we assess the impact of antipatterns on model performance in a setting where other antipatterns are cleaned out. Understanding whether antipatterns interact is crucial to assess the severity of each antipattern and prioritize our cleaning efforts accordingly. 

  Antipatterns like tailed distributions~\cite{hynes2017data}, distribution shift~\cite{hynes2017data}, or feature-row imbalance~\cite{Debie} exist in general research, but have not been adequately explored in SDP, despite the widespread use of ML in this field. Similarly, the lack of a formal taxonomy could have caused omissions in addressing antipatterns in other domains as well. We address this challenge by developing a taxonomy from the general literature on data quality, and identifying eight ML-specific data quality antipatterns, i.e., \textit{Schema Violations, Data Miscoding, Inconsistent Representation, Data Distribution Antipatterns, Packaging Antipatterns, Row-Feature Imabalance, Label Antipatterns}, and \textit{Correlation \& Redundancy}, and 14 antipattern sub-types. This taxonomy provides a structured understanding of the potential issues that could affect data quality, not only in software analytics, but in other fields as well. We then study the following research questions (RQs) in our case study:
\begin{itemize}
    \item \textbf{\rqone} \\ 
    \textit{Results:} SDP data contains \textit{schema violations, correlated \& redundant features, distribution antipatterns (tailed and not normal distributions), mislabeling,} and \textit{duplicated data} antipatterns. We found no instances of \textit{data miscoding, missing values} or the \textit{data drift} antipattern in the studied data. 
    Antipatters also overlap within projects and co-exist both on the row and column level in the tabular data. On the row level, 94\% of the rows with more than one antipattern include \textit{mislabels} and \textit{class overlap}. On the column-level, 90\% of the columns having two antipatterns include correlated \& redundant features and tailed features.
    
    \item \textbf{\rqtwoone} \\ 
    \textit{Results:} The order of cleaning impacts model performance, with certain orders having more pronounced effects than others. Removing class overlaps before  transformation consistently boosts the performance across all learners, emphasizing the importance of performing transformation at the end. In contrast, sequences like filtering before transformation and mislabel correction before overlap removal show varied impacts depending on the learner; simplistic classifiers like  Logistic Regression significantly benefit from mislabel correction before class overlap correction, while DNN shows robustness in its capacity to be resilient to different orders of cleaning antipatterns.

    \item \textbf{\rqtwotwo} \\
    \textit{Results:} 
    Although introduction of five antipatterns (\textit{Tailed, Mislabel, Class Overlap, Class Imbalance, Correlation \& Redundancy}) has a statistically significant impact on the performance metrics, most antipatterns only introduce a negligible to small effect on model performance. Among the antipatterns having an impact, only the \textit{Tailed} and \textit{Class Overlap} antipatterns result in a medium or large effect on some of the performance metrics. Overall there is heavy variability in the impact of different antipatterns across different learners.
    
    \item \textbf{\rqthree} \\ 
    Models constructed from data with different antipatterns exhibit moderate consistency in interpretation. This enhances the body of research on model interpretation which previously identified factors that impact the interpretation results, such as learner characteristics, hyperparameter tuning, and sampling randomness.
    
\end{itemize}


The remainder of the paper is organized as follows. Section~\ref{sec.lit} presents the relevant literature on data quality. Section~\ref{sec.back} explains the prestudy that summarizes the related work into a taxonomy of data quality antipatterns. Section~\ref{sec.casestudy} provides an overview of the case study setup along with a description of the case study parameters and various statistical tests used in this study. Section~\ref{sec.results} discusses the motivation, approach, and results for each of our research questions. Section~\ref{sec.threats} discusses threats to the validity of our study, while Section~\ref{sec.discussion} provides a discussion on our results. Section~\ref{sec.implications} provides the implications of our findings for researchers and practitioners, and finally, Section~\ref{sec.conclusion} concludes the paper.

%% file: LiteratureReview.tex
\section{Literature Review}\label{sec.lit}

While Section~\ref{sec.back} analyzes literature on data quality \textit{antipatterns} to develop a taxonomy, this section discusses research on 1) data quality issues in general domains, 2) data quality issues in software engineering and SDP, and 3) data quality tools.

    \subsection{Data quality issues in general.} The challenges in data quality for ML were identified as early as 1993~\cite{cortes1994limits} for different domains. For instance, Gupta et al.\cite{gupta2022data} analyzed data quality issues like inactive customers in the observation window and anomalous transaction behaviors that led to sub-optimal ML models used in banking systems. Kerr et al.\cite{kerr2007data} presented a case study of data quality and its critical nature in the healthcare industry. For environment monitoring domain, Okafor et al.~\cite{okafor2020improving} developed an ML framework to monitor the quality of data for low-cost IoT Sensors. Our research builds upon these works by investigating data quality  specifically in the SDP domain,  allowing us to uniquely identify the challenges and opportunities in addressing data quality attributes for software engineering applications, thereby contributing novel insights to the broader data quality literature. 
    
    Along with characterizing data quality issues in different domains, several studies have proposed \textit{tools for monitoring the quality of data.} For example, Grosso et al.\cite{grosso2023fast} developed an ML-based tool for real-time monitoring of particle detector data, and Ehrlinger et al.\cite{ehrlinger2019daql} proposed a DaQl (DAta Quality Library) as a generally-applicable tool to monitor the quality of data to increase the prediction accuracy of ML models. Similarly, Dai et al.\cite{dai2018improving} presented a framework built on deep learning and statistical models to identify data quality, and Shrivastava et al.\cite{shrivastava2020dqlearn} presented a toolkit for detecting, correcting, and explaining data quality for structured data. Our research aids such toolsmiths by providing a comprehensive taxonomy of data quality issues, helping them identify missed opportunities in the identification and mitigation of antipatterns in their domains.
    
    \subsection{Data Quality in Software Analytics.} In the realm of  defect prediction, several research works have meticulously curated and refined predictive features, underscoring the importance of data quality within this domain. Seminal works by Kamei et al.~\cite{kamei2012large} in 2013 initiated a shift towards identifying defect-prone (“risky”) software changes just-in-time, rather than focusing on broader entities like files or packages. Concurrently, Zimmermann et al.~\cite{zimmermann2009cross} pioneered the concept of Cross-Project Defect Prediction (CPDP), leveraging defect labels from one project to improve predictions in another, and outlined crucial features and machine learning models suitable for CPDP. Such seminal studies have spurred extensive subsequent research in the field, which are encapsulated in comprehensive survey papers such as those by Punitha et al.~\cite{punitha2013software}, Zhao et al.~\cite{zhao2023systematic}, and Thota et al.~\cite{thota2020survey}.   
    
    For general software engineering, research around data quality has increasingly focused on data management and operational best practices. For instance, in 2020, Jain et al.\cite{jain2020overview} presented an overview of data quality for machine learning applications at IBM, highlighting the importance of intelligently designed metrics in upholding data quality. The authors examined quality for structured data, unstructured text, and human in the loop validation by subject matter experts.
    Renggli et al.\cite{renggli2021data} from Microsoft Research presented a data quality view of MLOps, specifically providing guidelines and perspectives on designing an efficient MLOps pipeline. Abdallah et al.~\cite{abdallah2019big} uncovered data quality issues for big data, presenting factors from different perspectives such as management, user, processing, service pipeline, and data perspective. Moran et al.~\cite{moran2022important} assessed the impact of data quality by performing specific data preprocessings, and found that the significance of selecting a classifier decreases after applying an appropriate preprocessing step. Such aspects of data quality have been studied in research; however, our research is the first to empirically analyze the impact of multiple data quality antipatterns in the domain of SDP. Moreover, to the best of our knowledge, no software analytics study has yet focused on the impact of antipatterns on model interpretation. 

    Different software engineering domains can exhibit different ML engineering characteristics. For example, Ouatiti et al.~\cite{ouatiti2024impact} found that log level prediction models demonstrate very weak correlation in interpretation ranks with even contradictory model interpretations, a phenomenon not observed in SDP~\cite{lyu2021towards}). As such, software engineering researchers should evaluate antipatterns within the specific context of their respective domains.
    

%% file: taxonomy.tex
\section{Taxonomy of ML-specific Data Quality Antipatterns}\label{sec.back}
Although previous research has demonstrated the prevalence and impact of various data quality antipatterns, there is currently no comprehensive taxonomy of antipatterns in general literature. In this section, we perform a literature review and summarize data quality into a set of eight broad antipattern types, i.e., \textit{Packaging Antipatterns, Schema Violations, Data Miscoding, Feature-Row Imbalance, Data Distribution Antipatterns, Inconsistent Representation, Correlation \& Redundancy}, and \textit{Label Antipatterns}, which comprise of 14 sub-types. As the first step, this paper focuses on structured data quality antipatterns, in the context of ML. Future work should examine antipatterns for other data types (e.g., images, audio or video). Our taxonomy of antipatterns and strategies to address them will help researchers and practitioners in effective dealing of these antipatterns for their use cases. 

\textbf{\textit{Approach.}}
Given the absence of comprehensive studies listing these antipatterns, we focused on both tool-based solutions along with a traditional academic survey. 
Overall, we derived our taxonomy of antipatterns, by targeting three aspects: 

\begin{itemize}
    \item \textbf{Data Cleaning Tools}
    We began our search by looking for tools and identified specific antipatterns targeted by each tool. Particularly, we looked for data cleansing, profiling, and validation tools offering detection and mitigation solutions for different antipatterns. In Subsection~\ref{subsec.mitigation}, we present and cite various data profiling tools along with our findings on their  detection and mitigation strategies. 

    \item \textbf{Leveraging academic publication accompanying open source tools for an initial round of snowballing.} Often, open source tools are accompanied by a corresponding academic publication, which we used for snowballing (i.e., looking at the cited publications from that paper). For instance, tools like data linter\footnote{\url{https://github.com/brain-research/data-linter}} are accompanied by a scholarly paper~\cite{hynes2017data}, served as a foundational reference for further academic exploration. The data linter tool cleans 1) miscoding errors, 2) outliers and scaling issues, and 3) packaging errors, all three of which were directly incorporated into our taxonomy. Andrew NG\footnote{\url{https://en.wikipedia.org/wiki/Andrew_Ng}}'s~\cite{andrewNG} emphasis on data-centered innovations and development in ML~\footnote{\url{https://www.youtube.com/watch?v=TU6u_T-s68Y&ab_channel=SnorkelAI}} is centered around the inconsistent labeling antipattern. Akin to the data linter tool,  the TFDV (Tensorflow Data Validation) tool\footnote{\url{https://github.com/tensorflow/data-validation}} is accompanied by a publication by Breck et al.~\cite{breck2019data} and also served as a foundation reference for our academic exploration. TFDV detects antipattens like ``schema validation'', ``training serving skew'', and ``package errors'', which we directly adopted in our taxonomy. Tools like \url{DataPrep.ai} validate and clean entities (like Zip codes, phone numbers, and country names), a manifestation of Data miscoding. For each tool-academic based research, the second author employed a lightweight snowballing methodology, utilizing each tool's publication's references to expand our study. 

    \item \textbf{General academic search}.
    After the above two steps, the initial candidate list of publications was further refined and expanded upon by incorporating a systematic search using Google Scholar, employing terms such as ``data quality", ``data issues machine learning", and ``data quality artificial intelligence". This search yielded at least 50 relevant candidate publications along with their referenced publications via snowballing. Antipatterns discussed in these sources were cross-referenced with our existing taxonomy, and any new antipatterns were integrated accordingly. Notable additions in this phase included sub-types like class imbalance, feature-row imbalance, and concept drift.

\end{itemize}

Finally, the first two authors categorized the identified antipatterns into a taxonomy using a card sorting method. To ensure the robustness and coherence of the expanded taxonomy, the first two authors engaged in four collaborative sessions, each lasting approximately one hour, spread over several days. The categorization process was straightforward, with antipatterns logically grouped based on their characteristics. For instance, schema violations are clearly distinguished from label antipatterns, and the obvious placement of class overlap lies in the label antipatterns sub-type. The clarity of the antipattern characteristics facilitated a unanimous agreement on their categorization in the taxonomy.


\begin{figure}[!t]
      \makebox[\textwidth][c]
      {\includegraphics[width=1.4\columnwidth, keepaspectratio]{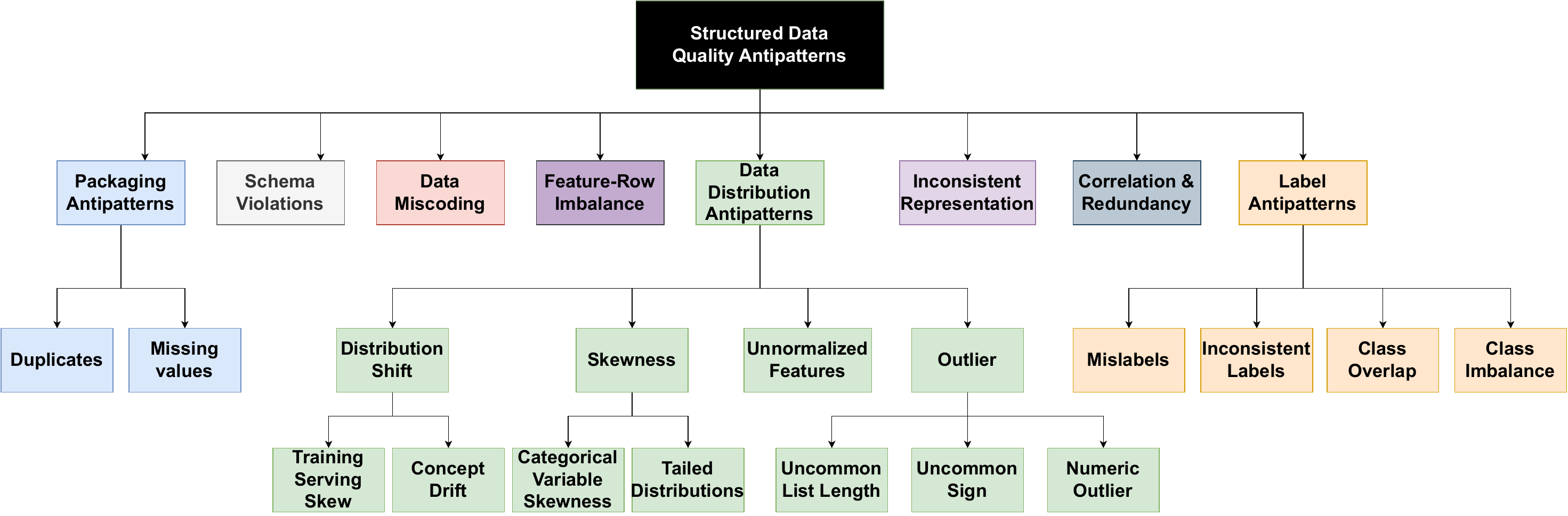}}
      \caption{A taxonomy of data quality antipatterns in structured data.}
      \label{fig.antipattern.taxonomy}
\end{figure}

In the subsections below, we discuss the 1) resulting taxonomy and 2) detection and mitigation strategies for each of the antipatterns from the resulting taxonomy.
 
\subsection{Resulting Taxonomy}\label{subsec.resulting_taxonomy}   

We identified a total of eight high-level types of antipatterns comprising a total of 14 sub-types. Figure~\ref{fig.antipattern.taxonomy} presents a taxonomy of the eight high-level data quality antipattern types along with their 14 sub-types. 
\begin{itemize}
    \item \textbf{Schema Violations.} 
    A data schema identifies a blueprint for the organization of a dataset based on a set of rules. Violations of the rules that explain the logic of the objects/associations/entities described by the data are schema violations. \\
    For relational  databases, integrity constraints are enforced to ensure the correctness of data within a defined schema. However, violations can occur if such rules are not enforced or are undefined. 
    Schema violations are quality antipatterns as data that violates its schema is likely to be inaccurate and therefore should introduce noise to the pattern that ML models are trying to learn.\\
    Schema rules have been proposed and used for data cleaning in the literature. For instance, in the context of defect prediction, Shepperd et al.~\cite{shepperd2013data}\textit{number\_lines} (total lines of in a source code file) $\geq$ \textit{number\_blank\_lines} (lines which have no code or comments, and are left blank) is a schema rule.

    
    \item\textbf{Data Miscoding.} Data miscoding refers to the incorrect assignment of data types to features, such as treating numerical data as strings. ML models are trained based on the incorrect data type, leading to suboptimal performance if, for instance, a numerical column is incorrectly treated as a string. Data miscoding encompasses the use of incorrect data types while performing ML operations during the training, validation, or serving phases of machine learning processes~\cite{amershi2019software}. For instance, in the training phase, models trained from an incorrect data type may learn improper information from the miscoded features.
    
    This antipattern can be dynamically introduced while performing data-wrangling tasks. For instance, an \textit{enum} can be incorrectly read as a \textit{real number} in memory. Hynes et al.~\cite{hynes2017data} explained other manifestations of data miscoding as \textit{``numbers as strings'', ``tokenizable strings'', ``circular domain as linear'', ``date/time as a string, zip code as a number'', and ``integer as float''}. 
    
    Automatically inferring the data schema using libraries can also cause miscoding, for instance, reading data from a CSV file in the $read\_csv$ function of python's pandas library enables automatic inferring of data types, which if not carefully managed with explicit typecasting can be a source of the ``data miscoding'' antipattern.
    
    Removing this antipattern is a vital common practice, and to combat this antipattern,  practitioners and researchers have to explicitly typecast data for building ML models. For instance, Sharma et al.~\cite{sharma2022performance} mentioned ``all the incorrect data types of the features changed to their required datatype'' while evaluating ML-based feature selection approaches for breast cancer detection.
    	
    \item\textbf{Inconsistent Representation.}
    Two or more ways of representing the same concept within the data. This antipattern can also manifest as an inconsistency with the lower and upper case for a string feature. For instance, in a dataset with a country code feature, countries like Canada could be represented by the string ``CAN'' in some entries while ``CANADA'', or lower case string ``canada'' or a camel case ``Canada'' in others leading to inconsistent representations.  In time series data, this antipattern could manifest as inconsistent units (e.g., cm and inches) for the same feature~\cite{gitzel2016data}. In the tabular data, this antipattern occurs at the row level, whereas the column-level representation of the same/similar entities could exhibit other antipatterns like correlation and redundant features, which we explain separately.

    \item\textbf{Data Distribution Antipatterns.}
    Several characteristics of  distributions have been considered as antipatterns by prior research~\cite{hynes2017data,breck2019data}. We observed seven sub-types of distribution errors from the literature. 
    \begin{itemize}
        \item \textbf{Unnormalized Features.\footnote{We use the definition of Unnormalized features from Hynes et al.~\cite{hynes2017data}. This is not the same as Not Normal (i.e., Not Gaussian) distributions.}}
        Distribution of one or more features that are widely different from the distribution of other features. 
        
        \item \textbf{Outliers.}
        Under the sub-category of outliers, we have further sub-categories based on the type of outlier, i.e., size, numerical outlier, or sign. We provide a definition for each below:

        \begin{itemize}
            \item \textbf{Tailed Features.} Extreme values that significantly affect the mean are called ``tailed''~\cite{hynes2017data}. Typically histograms need to be examined to ensure that the data follows expected distributions.

            \item \textbf{Uncommon list length.} For features composed of lists, some instances have a different length than the common or correct length~\cite{hynes2017data}.
        
            \item \textbf{Uncommon sign.} For numeric features, some values have a different sign (+/-) from the rest of the data~\cite{hynes2017data}. Particularly, the sign of one or more rows for a feature itself is an outlier.  
         
        \end{itemize}
        
        \item \textbf{Training serving distribution skew.} 
        A difference between the distribution of data used during the model training and deployment (a.k.a. serving) phase. When these data distributions are significantly different (i.e., there is a heavy ``skew''), the model can perform poorly on the new data~\cite{breck2019data}. Training-serving skew is considered an antipattern because if the data used to train the model does not accurately represent the data that the model will encounter in production, it may result in issues such as overfitting, bias, or poor performance.

        \item \textbf{Data drift.} Drift is a distribution change between different temporal snapshots of data, which can be problematic if models trained on differently distributed datasets enforce performance changes. Data drift could be prevalent in both independent and dependent features~\cite{webb2016characterizing,vzliobaite2016overview}.

        \textit{Covariate drift} occurs when there is a shift in the independent variables over time. For instance, a cats/dogs classifier trained on images of golden retrievers has a change in deployment, when at a different time, a newer breed of dog, say bulldog, is introduced. 
        
        \textit{Class drift a.k.a. prior probability drift} indicates a shift in the target variable. For the cats/dogs classifier, class drift could manifest as the addition of an entirely new class, say rabbit images.

        \item \textbf{Constant Features.} Features that have no variance or have the same value for all rows. These features do not add any value to the ML model, and hence, Constant features are an antipattern.  
        \end{itemize}

    \item\textbf{Packaging Antipatterns.}
    As mentioned by Hynes et al.~\cite{hynes2017data}, ``Packaging errors identify problems with the organization of the data''. These issues have the potential to seep into the dataset while packaging the data for building ML models, perhaps during the data preprocessing step of the ML pipeline~\cite{amershi2019software}. Packaging errors manifest as \textit{Empty} or \textit{Duplicated} values. 
    \begin{itemize}
        \item \textbf{Empty Values.} Missing or null values must be dealt with before using the data for ML tasks since ignoring or omitting missing values may result in biased or misinformed analysis~\cite{emmanuel2021survey}.
        \item \textbf{Duplicates}. Repeated entries for one or more features~\cite{ilyas2019data,hynes2017data,shepperd2013data}. This antipattern occurs at the row-level as opposed to the Constant features antipattern, which occur at the column-level.
    \end{itemize}

    \item\textbf{Label Antipatterns.} Antipatterns that exist in the dependent variable of the training data. 
    \begin{itemize}
        \item \textbf{Inconsistent Labeling.} 
        Incorrect values assigned to a label due to different and inconsistent labeling strategies employed by the labelers. In the case of labels assigned via crowd-sourcing (i.e., organizations employ large teams to annotate ML datasets), inconsistencies in the labeling strategy employed by different teams can cause performance issues for ML models~\cite{andrewNG}. Similar issues can occur with non-human labeled datasets labeled with inconsistent strategies~\cite{ilyas2019data}.
        
        \item \textbf{Mislabels.} Response variable labeled via an incorrect labeling strategy. As opposed to \textit{inconsistent labeling}, where the labels are inconsistently assigned, mislabels include labels that are systematically assigned in an incorrect manner.
        
        For instance, in the case of SDP, Da Costa et al.~\cite{da2016framework} suggested improvements to label buggy instances by enhancing the SZZ algorithm by encorporating the earliest bug appearance, future impact of changes, and realism of bug introduction. Indeed, their labeling strategies lead to performance improvements for SDP models, as empirically shown by Yatish et al.~\cite{yatish2019mining}.
        
        \item \textbf{Class Overlap.} Instances having similarities in feature values although they belong to different classes, which is a substantial obstacle to their correct classification by a ML model~\cite{vuttipittayamongkol2021class}. These instances have more than one class sharing a common region in the data space. The adverse effect of Class overlap towards an ML model's performance often goes hand in hand with Class imbalance~\cite{chen2018tackling,denil2010overlap,vuttipittayamongkol2021class,gong2019empirical}.

        \item \textbf{Class Imbalance.} This antipattern occurs when the majority of instances are labeled with one class, whereas very few instances are labeled with the other class(es)~\cite{guo2008class}. Class imbalance can significantly impact the performance of ML classifiers~\cite{johnson2019survey,yang2020rethinking,ali2019imbalance}.
    \end{itemize}

    \item\textbf{Correlation \& Redundancy.}
    We identify \textit{correlation \& redundancy} as one antipattern group since their presence while training ML models renders the model's interpretation inconsistent, as identified by Jiarapakdee et al.~\cite{jiarpakdee2019impact} for software analytics.
    Following Jiarpakdee's seminal work in 2019, future defect prediction research (e.g., \cite{rajbahadur2021impact,harzevili2021analysis,gong2019empirical}) has removed correlated and redundant features before performing ML tasks. An explanation of \textit{correlation \& redundancy} is provided below:
    \begin{itemize}
        \item \textbf{Correlation.} A relation between two or more variables. To avoid multicollinearity, a situation where two or more predictor variables in a regression model are correlated, leading to unreliable estimates, it is crucial to remove correlated variables before ML model training ~\cite{daoud2017multicollinearity}.
        
        \item  \textbf{Redundancy.} One or more independent variables being represented by other independent variables. Redundant variables interfere with each other, distorting the modeling relationship between the independent and the dependent variables. Hence, such variables need to be removed before training the ML model~\cite{mcintosh2016empirical}. 
    \end{itemize}

    \item \textbf{Feature-row imbalance} 
    For training ML models, the data size should not be imbalanced, where too many features w.r.t. the number of rows or vice-versa can cause model performance issues. We explain both these cases below:
    \begin{itemize}
        \item \textbf{Over-dimensionality} can cause issues that are commonly attributed to the ``Curse of Dimensionality", i.e., the number of dimensions (features) being much larger than the number of observations (samples/rows). Prior research~\cite{Debie,vong2019additional,pudjihartono2022review} has attributed issues such as poor classification accuracy, overfitting, decreased model interpretability, and increased computational time/resources to the curse of dimensionality.

        \item \textbf{Under-dimensionality} is a feature-row imbalance where there are relatively more rows than features (dimensions). Such situations cause ML model underfitting, leading to high model bias, and potential data redundancy (i.e., with too many rows, patterns usually start to be repetitive~\cite{hastie2009elements,l2017machine}).
        
    \end{itemize}    
    
\end{itemize} 

Our taxonomy of ML-specific data quality antipatterns is the first step towards a foundational framework for understanding the landscape of data quality issues in software analytics. This taxonomy not only categorizes the prevalent antipatterns but also raises critical questions about their broader implications in ML model development and performance as reflected in the research questions of this study.

%% file: Appendix_mitigation_strategies.tex
\subsection{Antipattern Detection and Mitigation Strategies}\label{subsec.mitigation}

In this sub-section, we describe the \textbf{detection} approach and \textbf{mitigation} process for each of these eight top level and 16 sub-categories of antipatterns discussed. 

    \subsection{Packaging Antipatterns}
        \begin{enumerate}
        \item \textit{Missing values}\\
            \textbf{Detection Strategies:} Missing values can be detected by checking\footnote{\url{https://pandas.pydata.org/docs/reference/api/pandas.isnull.html}} for null data.\\
            \textbf{Mitigation Strategies:} Dropping or imputing the missing values. Typical imputation strategies include Hot Deck, Cold Deck, Mean Substitution, Non-negative matrix factorization, and regression~\cite{ilyas2019data}.
        
        \item \textit{Duplicates} \\
            \textbf{Detection Strategies:}
            \begin{itemize}
            \item	Detecting\footnote{\url{https://pandas.pydata.org/docs/reference/api/pandas.DataFrame.duplicated.html}} exact duplicates across rows.
            \item	Using distance metrics for detecting numeric duplicates~\cite{ilyas2019data}.
            \item	String duplicates can be checked using a) character-based b) token-based, and c) phonetics-based similarity metrics~\cite{ilyas2019data}.
            \end{itemize}
            \textbf{Mitigation Strategies:}
                \begin{itemize}
                \item Dropping\footnote{\url{https://pandas.pydata.org/docs/reference/api/pandas.DataFrame.drop_duplicates.html}} duplicates.
                \item Clustering duplicates and record fusion. Hassanzadeh et al.~\cite{hassanzadeh2009framework} provide a framework for evaluating clustering algorithms for data duplication. Clustering is followed by fusing records, which can be done by probabilistic methods~\cite{ilyas2019data} or other conflict resolution strategies provided by~\cite{bleiholder2009data}.
                \item Using data duplication tools: Academic open-source tools include AJAX~\cite{galhardas2000ajax}, Febrl~\cite{christen2008febrl}, while commercial tools include Data Tamer~\cite{stonebraker2013data}.
                \item Human-involved de-duplication: Using crowdcourcing platforms like Amazon Turk, Corleone~
\cite{gokhale2014corleone}. Other tools like CrowdER~\cite{wang2012crowder} provide hybrid (human-machine) workflow.
                \end{itemize}
        \end{enumerate}

\subsection{Schema Violations}
    \textbf{Detection Strategies:} Rule based unit testing (i.e., checking the data values against a set of pre-defined rules). Tools like \cite{grex} can be used to encode schema rules and check for violations.\\
    \textbf{Mitigation Strategies:} Either removing or imputing the violations. Imputation strategies are discussed in the previous subsection. 

\subsection{Data Miscoding}
    \textbf{Detection Strategies.} Explicit type checking\footnote{\cite{caveness2020tensorflow,hynes2017data}, can perform this.}.\\
    \textbf{Mitigation Strategies.} In the scripts for data preprocessing for model training and serving, explicit type casting\footnote{\url{https://pandas.pydata.org/docs/reference/api/pandas.DataFrame.astype.html}} can help prevent this antipattern.

\subsection{Row Feature Imbalance}
    \textbf{Detection Strategies.} Checking the ratio of the number of rows with respect to the number of columns.\\
    \textbf{Mitigation Strategies} include downsampling or upsampling. Downsampling methods include:
    \begin{itemize}
        \item \textbf{Feature selection} can be done by wrapper methods~\cite{hall1999correlation} or filter methods~\cite{guyon2003introduction}.
        \item \textbf{Dimensionality reduction} can be done by Principal Component Analysis (PCA), Linear Distriminant Analysis (LDA), etc. algorithms~\cite{ilyas2019data}.
        \item \textbf{Clustering} can be done by K-Means, Hierarchical Clustering, etc. algorithms~\cite{ilyas2019data}.
    \end{itemize}

    Upsampling or data augmentation can be done via algorithms like Synthetic Minority Over-Sampling technique (SMOTE), Adaptive Synthetic Sampling (ADASYN), etc.~\cite{ilyas2019data}

\subsection{Data Distribution Antipatterns} 

    \subsubsection{Distribution Shift}

    \begin{itemize}
        \item \textbf{Training-serving skew\\}
         \textbf{Detection}: Checking data drift using L-infinity distance for categorical features, and Jensen-Shannon divergence for numeric features\footnote{TFX~\cite{breck2019data} and TFDV~\cite{hynes2017data} provides this capability}.\\
        \textbf{Mitigation strategies}: Reshuffling or taking alternative batches of training and serving data.
    
        \item \textbf{Concept Drift\\}
        \textbf{Detection Strategies:} such as 1) \textit{Sequential analysis}~\cite{muthukrishnan2007sequential,ikonomovska2011learning}, 2) Control charts~\cite{gomes2011learning,ross2012exponentially}, and 3) Contextual analysis~\cite{harries1998extracting,klinkenberg2004learning}. Gama et al.~\cite{gama2014survey} provide a comprehensive survey explaining these three strategies.\\
        \textbf{Mitigation Strategies:} Online learning, periodically retraining the data, feature dropping\footnote{\url{https://neptune.ai/blog/concept-drift-best-practices}}.
    \end{itemize}

    \subsubsection{Outliers}
    \textbf{Detection Strategies:}
    \begin{itemize}
        \item \textit{ML-based detection.} ML outlier detection algorithms include Seq2Seq, Isolation Forest, and Variational Auto-Encoders\footnote{Alibi (\url{https://github.com/SeldonIO/alibi-detect}) detect tool provides implementations}.
        \item \textit{Distance-based detection.~\cite{ilyas2019data}}  Global or local methods depending on the reference population used when determining whether a point is an outlier.
        \item \textit{Statistics-based.~\cite{ilyas2019data}} Hypothesis testing methods like Grubbs Test~\cite{grubbs1950sample} and Tietjen-Moore Test~\cite{tietjen1972some}.
        \item \textit{Fitting distribution methods:} Fitting a distribution or inferring a probability density function based on the observed data, for example, local distance-based outlier detection method.~\cite{tao2006mining}
    \end{itemize}
    \textbf{Mitigation Strategies:} Dropping the outliers~\cite{laurikkala2000informal}, or transforming the feature having outliers.
    
    \subsubsection{Skewness}
    \begin{enumerate}
        \item \textit{Categorical variable skewness}\\
        \textbf{Detection Strategies:} Evaluating the frequency distribution (same as class imbalance but for explanatory variables).\\
        \textbf{Mitigation Strategies:} Using over-sampling or under-sampling algorithms to balance the affected variables~\cite{chawla2009data}.
        \begin{itemize}
            \item Oversampling techniques like random oversampling, SMOTE, ADASYN, and augmentation.
            \item Undersampling techniques like random, cluster, and Tomek links algorithms.
        \end{itemize}

        \item \textit{Tailed Distributions}\\
        \textbf{Detection Strategies:} 1) tools like~\cite{grex,breck2019data} enable feature inspection. 2) Hynes et al.'s Data Linter tool~\cite{hynes2017data} enables automatic detection of tailed features.\\
         \textbf{Mitigation Strategies}: Using Scalars, Transformers, and Normalizers (e.g., min-max scalar, z-score normalizer, quantile transformer, log transformer).\\
    \end{enumerate}

    \subsection{Unnormalized Features} 
    \textbf{Detection Strategies:} Inspecting feature distributions using tools like ~\cite{grex,caveness2020tensorflow,hynes2017data}.\\
    \textbf{Mitigation Strategies:} Transformations like min-max scaling, Z-scoring~\cite{hynes2017data} robust scaling, etc. 

\subsection{Inconsistent representation}\label{subsec.incon}
    \textbf{Defection Strategies: } Rule based testing of inconsistencies. This can be done by: 1) Using regex for verifying patterns\footnote{\cite{dataprep} can automatically perform this}.2) Validating categorical values to belong within a predetermined set\footnote{\cite{caveness2020tensorflow,grex} can perform this}.\\
    \textbf{Mitigation Strategies}
    1) Removing inconsistencies. 2) Rule-Based replacing inconsistencies can be done by unifying representations programmatically or using external tools.\footnote{\cite{dataprep} provides capability to validate and clean features like \textit{country name, email address, geographic coordinates, IP address, phone number, URLs, postal codes, url, social security identifiers, VAT, invoice, and business identifiers,} etc.}

\subsection{Label Antipatterns}
	\subsubsection{Inconsistent labeling}
    Refer to Subsection~\ref{subsec.incon} for \textbf{detection} and \textbf{mitigation} strategies.
     
    \subsubsection{Mislabels}
	\textbf{Mitigation strategies: }
    \begin{itemize}
        \item \textit{Human Verification:}
        \begin{itemize}
            \item Label verification using crowdsourcing~\cite{vaughan2017making} can correct labeling errors.
            \item Clearly defining labeling requirements and using skilled labelers.
            \item Using consensus.
            \item Using third party labeling and verification services like Turk\footnote{\url{https://www.mturk.com}}, Google reCaptcha\footnote{\url{https://www.mturk.com}}, and ByteBridge\footnote{\url{https://www.bytebt.com}}.
        \end{itemize}	

        \item \textit{Programmatic Labeling:} Instead of labeling data by hand, users write labeling functions, which programmatically label or abstain data points~\cite{ratner2020snorkel}.
    \end{itemize}

    \subsubsection{Class Overlap} 
   \textbf{Detection Strategies:} Detection strategies for Class overlap have been discussed in Section~\ref{sec.casestudy}.\\
	\textbf{Mitigation Strategies}: Separating, Merging or Discarding~\cite{das2013handling} class overlapping rows.

    \subsubsection{Class Imbalance}
   \textbf{Detection Strategies:} Detection strategies include checking the ratio of the different classes in the training data, as discussed in Section~\ref{sec.casestudy}.\\
    \textbf{Mitigation strategies}
    \begin{itemize}
        \item \textit{Data‑sampling methods} like Random Over-Sampling (ROS), Random Under Sampling (RUS), synthetic minority over sampling technique (SMOTE)~\cite{rodriguez2014preliminary}.
        \item \textit{Feature selection methods} can identify key features that distinguish between classes, improving classifier accuracy in uneven datasets~\cite{grobelnik1999feature}. A survey by Leevy et al~\cite{leevy2018survey} provides a detailed explanation different feature selection strategies.
        \item \textit{Algorithm-level methods} like Chi-FRBCS-BigDataCS algorithm~\cite{lopez2015cost}, a Fuzzy Rule-Based Classification System (FRBCS), linguistic cost-sensitive FRBCS~\cite{chi1996fuzzy} are designed to manage the challenges involved with class imbalance~\cite{leevy2018survey}.
    \end{itemize}

\subsection{Correlation \& Redundancy}
\textbf{Detection Strategies:} 
\begin{itemize}
    \item \textit{Correlation.} For detection of correlation between 1) numerical variables, Pearson or Spearman correlation metrics can be used; 2) Numerical and categorical using ANOVA or Kruskal-Wallis H test; 3) Categorical variables using Chi-Squared test.
    \item \textit{Redundancy} analysis can be done by 1) building parametric additive model which determines how well a variable is predicted from the remaining explanatory variables, and dropping the most predictable variable in a step wise fashion~\footnote{\url{https://cran.r-project.org/web/packages/Hmisc/index.html}}.

    This approach is opposed to the 2) ordination approach, employed by SciKit\footnote{\url{http://scikit-bio.org/docs/0.5.4/generated/generated/skbio.stats.ordination.rda.html}}. In this approach, the explanatory variables (\texttt{x}) are fitted to the explained variable (\texttt{y}) using techniques like redundancy analysis (RDA), followed by a Principal Component Analysis (PCA) on the fitted values to identify and remove redundant variables.

\end{itemize}
\textbf{Mitigation Strategies:} Removal of the correlated and redundant features.\\

%% file: case_study.tex
\section{Case Study Setup}\label{sec.casestudy}

In this section, we describe the case study setup that is used to answer the research questions. 
\begin{table}[!h]
\caption{Metrics from the studied SDP dataset obtained from Yatish et al.~\cite{yatish2019mining}}
\label{tab.data}
\begin{tabular}{l|l}
\textbf{Metric Type} & \textbf{Metric Names} 

\\ \hline
Code Metrics &
  \begin{tabular}[c]{@{}l@{}}
  AvgCyclomatic, AvgCyclomaticModified, \\
  AvgCyclomaticStrict, AvgEssential, \\
  AvgLineBlank, CountDeclClassMethod,\\
  AvgLineCode, Avg-LineComment,\\ 
   AvgLine, CountDeclClassVariable, CountStmt, \\
  CountDeclFunction, CountDeclInstanceMethod, \\ 
  CountDeclInstanceVariable, CountDeclMethod, \\
  CountDeclMethodDefault, CountLine,\\
  CountDeclMethodPrivate, CountLineBlank,\\ 
  CountDeclMethodProtected, CountDeclClass, \\
  CountDeclMethodPublic, CountLineCode, \\
  CountLineCodeDecl, CountLineCodeExe, \\ 
  CountLineComment, CountSemicolon, \\
  CountStmtDecl, CountStmtExe, MaxCyclomatic, \\
  MaxCyclomaticModified, MaxCyclomatic-Strict, \\
  RatioCommentToCode, SumCyclomatic, \\
  SumCyclomaticModified, SumCyclomaticStrict, \\
  SumEssential, CountClassBase, \\
  CountClassDerived, MaxInheritanceTree, \\
  PercentLackOfCohesion, CountClassCoupled, \\
  CountInput \{Min, Mean, Max\},\\ 
  CountOutput \{Min, Mean, Max\}, \\
  CountPath \{Min, Mean, Max\}, \\
  MaxNesting \{Min, Mean, Max\}
  \end{tabular}  \\ \hline
Process Metrics      & \begin{tabular}[c]{@{}l@{}}COMM, ADDED\_LINES, DEL\_LINES, \\ ADEV, DDEV\end{tabular}\\ \hline
Ownership Metrics    & \begin{tabular}[c]{@{}l@{}}MINOR\_COMMIT, MINOR\_LINE, \\ MAJOR\_COMMIT, MAJOR\_LINE, \\ OWN\_COMMIT, OWN\_LINE\end{tabular}\\ \hline
\end{tabular}
\end{table}

\subsection{Studied Dataset}

\begin{table}[b!]
\caption{The studied project data, version count and the data size.}\label{tab:project_data}
\centering
\begin{tabular}{|l|c|r|}
\hline
\textbf{Project Name} & \textbf{\#Versions} & \textbf{Median \#Rows} \\
\hline
Activemq & 5 & 2040 \\
Camel & 2 & 7517 \\
Derby & 3 & 2206 \\
Groovy & 3 & 821 \\
Hbase & 3 & 1669 \\
Hive & 3 & 1560 \\
Jruby & 4 & 1054.5 \\
Lucene & 4 & 1325.5 \\
Wicket & 3 & 1763 \\
\hline
\end{tabular}
\end{table}
Software defect prediction (SDP) is the process of using software-related metrics as explanatory and response variables towards building ML or statistical models for identifying and forecasting potential defects in software applications before they occur. Such models help to identify patterns and correlations that can predict future software bugs, improving the quality of future releases. 

In this study, we use the SDP dataset from Yatish et al.~\cite{yatish2019mining}. The metrics used in this dataset are code related (i.e., metrics that define the relationship between code properties and software quality), owner related (i.e., metrics that describe the relationship between module ownership and software quality) and process related (i.e., metrics that define relationship between development activities and software quality)~\cite{yatish2019mining}.
Table~\ref{tab.data} presents the 65 metrics adapted from Yatish et al. used in this study. Yatish et al.'s study delved into the characteristics of the response label in SDP, can be interpreted in our study context as an effort towards ``mislabel correction". The dataset comprises nine  popular open source libraries and their 30 project versions, as aggregated in Table~\ref{tab:project_data}.


\subsection{Antipattern Detection Techniques}
In this sub-section, we outline the procedure for checking the different data quality antipatterns within our studied SDP dataset.
We describe the approach to detect each of these data quality antipatterns below.

\begin{table*}[htbp]
        \caption{
        Schema Violation Rules. Rules marked with an asterisk($^{*}$) are SDP domain-specific.  
        }  \label{tab.schema}

        \hspace*{-0.91cm}\begin{tabular}{c|p{10.5cm}|p{5.5cm}}
        
        \textbf{Rule} & \textbf{Rule Description} & \textbf{Rule Pseudo-Code} \\ \hline
        
        R1$^{*}$ & All metrics should be greater or equal to zero & Metric \textgreater{}= 0 \\ \hline
        
        R2$^{*}$ & \#physical lines should be greater than or equal to the \#lines containing comments. & CountLine \textgreater{}= CountLineComment \\ \hline
        
        R3$^{*}$ & \#physical lines should be greater than or equal to the average lines of code. & CountLine \textgreater{}= AvgLine \\ \hline
        
        R4$^{*}$ & \#physical lines should be greater than or equal to the \#declarative lines of code. & CountLine \textgreater{}= CountLineCodeDecl \\ \hline
        
        R5$^{*}$ & \#physical lines should be greater than or equal to the \#executable lines of code. & CountLine \textgreater{}= CountLineCodeExe \\ \hline
        
        R6$^{*}$ & \#physical lines should be greater than or equal to the sum of declarative and executable lines of code. & CountLine \textgreater{}= CountStmt \\ \hline
        
        R7$^{*}$ & \#physical lines should be greater than or equal to the \#blank lines of code. & CountLine \textgreater{}= CountLineBlank \\ \hline
        
        R8$^{*}$ & \begin{tabular}[c]{@{}l@{}}\#physical lines should be greater than or equal to the sum of their components,\\i.e., sum of blank lines, comment lines, declarative and executable lines.\end{tabular} & \begin{tabular}[c]{@{}l@{}}CountLine \textgreater{}= \\
        (CountLineBlank + CountLineComment \\ + CountLineDecl + CountLineExe)\end{tabular} \\ \hline
        
        R9$^{*}$ & \#statements should be the sum of declarative statements and executable statements. & CountStmt = CountLineCodeDecl + CountLineCodeExe \\ \hline
        
        R10$^{*}$ & The ratio of comment to code should be the \#comment lines divided by the total \#code lines. & RatioCommentToCode = CommentLineCode / CountLineCode \\ \hline
        
        R11$^{*}$ & For all nested functions or methods, average physical lines should be greater than the average lines containing code. & AvgLine \textgreater{}= AvgLineCode \\ \hline
        
        R12$^{*}$ & For all nested functions or methods, average physical lines should be greater than the average lines containing comments. & AvgLine \textgreater{}= AvgLineComment \\ \hline
        
        R13$^{*}$ & For all nested functions or methods, average physical lines should be greater than the average lines containing blank lines. & AvgLine \textgreater{}= AvgLineBlank \\ \hline
        
        R14$^{*}$ & \begin{tabular}[c]{@{}l@{}}For all nested functions or methods, average physical lines should be greater than\\the sum of  average lines containing code, comments, and blank lines.\end{tabular} & AvgLine \textgreater{}= (AvgLineCode + AvgLineComment + AvgLineBlank) \\ \hline
        
        R15 & \begin{tabular}[c]{@{}l@{}}For all nested functions or methods, the sum of cyclomatic complexity\\  should be greater than the maximum cyclomatic complexity.\end{tabular} & SumCyclomatic \textgreater{}= MaxCyclomatic \\ \hline
        
        R16 & \begin{tabular}[c]{@{}l@{}}For all nested functions or methods, sum of strict cyclomatic complexity \\ should be greater than the maximum strict cyclomatic complexity.\end{tabular} & SumCyclomaticStrict \textgreater{}= MaxCyclomaticStrict \\ \hline
        
        R17 & \begin{tabular}[c]{@{}l@{}}For all nested functions or methods, sum of modified cyclomatic complexity \\ should be greater than the maximum modified cyclomatic complexity.\end{tabular} & SumCyclomaticModified \textgreater{}= MaxCyclomaticModified \\ \hline
        
        R18 & \begin{tabular}[c]{@{}l@{}}For all nested functions or methods, maximum of cyclomatic complexity \\ should be greater than the average of cyclomatic complexity.\end{tabular} & MaxCyclomatic \textgreater{}= AvgCyclomatic \\ \hline
        
        R19 & \begin{tabular}[c]{@{}l@{}}For all nested functions or methods, maximum of strict cyclomatic complexity \\ should be greater than the average of strict cyclomatic complexity.\end{tabular} & MaxCyclomaticStrict \textgreater{}= AvgCyclomaticStrict \\ \hline
        
        R20 & \begin{tabular}[c]{@{}l@{}}For all nested functions or methods, maximum of modified cyclomatic complexity \\ should be greater than the average of modified cyclomatic complexity.\end{tabular} & MaxCyclomaticModified \textgreater{}= AvgCyclomaticModified \\ \hline
\end{tabular} 
\end{table*}

\subsubsection{Schema Violations}
    A schema refers to predefined rules and structures that dictate the acceptable values and relationships for data. These rules help identify and rectify implausible or erroneous data points. 
    Since there is no formal definition of SDP schema rules in academia, we first formulate a schema for our studied SDP data, then check for deviance from the logic defined within these rules. 
    Our approach for finding schema rules is two-phased. Firstly, we referred to prior literature to check the rules and associations between the SDP metrics. For instance,  Shepperd et al.~\cite{shepperd2013data} defined referential integrity checks (like \texttt{$\#LINES \geq \#CODE + \#COMMENT$}) for the NASA dataset. 
    
    Five such rules by Shepperd et al. were applicable to our dataset (\texttt{R3-R7} in Table~\ref{tab.schema}), hence we reused them.
    Next, we analyzed the definitions for each metric defined in Table~\ref{tab.data} to develop more rules. 
    Particularly, we used general knowledge that the i) sum of a metric is greater or equal to the max value, and ii) maximum should be greater or equal to the average value, to generate additional rules.

    Table~\ref{tab.schema} presents the set of 20 identified rules with a unique identifier for every rule. \texttt{R1-R14} are domain-specific (SDP domain) rules and are marked with $^{*}$, whereas, the remaining rules (i.e., \texttt{R15-20}) are based on common statistics. In the table, rules like \texttt{R9}, indicating that the ``\#total statements'' should be the sum of ``declarative'' and ``executable'' statements, are based on the metric definitions obtained from Scitools\footnote{\url{https://support.scitools.com/support/solutions/articles/70000582223-what-metrics-does-understand-have-}}. In particular, \textit{declarative} statements include variable assignment (e.g., \texttt{x=10;}), while executable statements include method calls (e.g., \texttt{System.out.println("Hello, World!");}). On the contrary, rules like \texttt{R18} are based on the statistical rule that the maximum value should be greater than the average value.

    \subsubsection{Distribution Antipatterns}
    We use Hynes et al.`s~\cite{hynes2017data} implementation to identify non-normalized features, and tailed features.
    \begin{itemize}
    \item \textbf{Constant Features.} We identify constant features by checking whether the variance of a feature is zero. Such an approach can only be adopted for numerical features. We do not have any categorical features within our studied SDP data, and hence, have not adopted any technique to remove any constant categorical feature.

    \item \textbf{Data Drift.} 
    For each project in our studied SDP dataset, we detect drift between its consecutive versions. For instance, to detect data drift for the project \textit{Derby} across versions, \textit{10.2.1.6,  10.3.1.4}, and \textit{10.5.1.1}, we study drift between i) \textit{10.2.1.6 and 10.3.1.4}, ii) \textit{10.3.1.4 and 10.5.1.1}. Similarly, we compared all consecutive version data for each of the nine projects in our dataset.
    Notably, our approach employs TFDV's (tensorflow data validation tools) implementation to detect this consecutive drift. 

    \item\textbf{Unnormalized Distributions.} This calculation involves trimmed means (the mean calculated after removing the top and bottom 10\% of data points, which may include extreme low or high values) and standard deviations for these features, thereby discounting extreme values at both ends of the spectrum. This approach focuses the analysis on more representative central data points. For each numeric feature, the detector assesses the deviation of its mean and standard deviation from these trimmed statistics using z-score calculations. Features whose mean or standard deviation deviances exceed a certain threshold\footnote{We use the default threshold values obtained from Hynes et al.~\cite{hynes2017data}} are considered unnormalized.
    
    \item\textbf{Tailed Distributions.} Along with deriving this antipattern from Hynes et al.~\cite{hynes2017data}, we derived the implementation from their research. Notably, the implementation relies on checking whether the deviation of a feature mean is greater than that of its trimmed mean. The implementation detects custom missing/placeholder values (e.g., -999), along with finding obvious statistical outliers.  
    \end{itemize}

    \subsubsection{Package Antipatterns}
    \begin{itemize}
    \item \textbf{Empty Values}. We use pandas' \texttt{isnull}\footnote{\url{https://pandas.pydata.org/docs/reference/api/pandas.isnull.html}} function to check for empty values.
    
    \item \textbf{Duplicates}. 
    In the studied SDP dataset, a row represents a source code file within a specific project version. If that file has identical feature values as another file, then such files (rows) are duplicated. 
    We detect duplicates on a row level using the \textit{duplicated()} function\footnote{\url{https://pandas.pydata.org/docs/reference/api/pandas.DataFrame.duplicated.html}} from the Pandas package. 
    
    \end{itemize}

    \subsubsection{Label Antipatterns}
    \begin{itemize}
        \item
	Yatish et al.'s~\cite{yatish2019mining} dataset consists of labels (i.e., whether a file is buggy) assigned based on two different labeling strategies. At the core of data preparation for SDP is the identification of post-release defects, where modules are modified to fix defects post-release. The traditional approach for identifying post-release defects involves extracting specific keywords and issue IDs from commit logs in the version control system using regular expressions~\cite{kamei2012large,d2012evaluating}. Labels derived from this heuristic-based approach are used to train ML models to predict the bugginess of future code.
 
    Yatish et al. refined this traditional approach following Da Costa et al.~\cite{da2016framework}, by considering affected releases in an issue-tracking system to identify defect-introducing commits. This realistic method incorporates release data when finding bug-introducing commits, unlike the traditional ``heuristic'' approach. We use the feature names ``RealBug'' and ``HeuBug'' from Yatish's dataset to label correct and mislabeled tuples.

    \item \textbf{Class Imbalance.} We determine the presence of class imbalance by evaluating the ratio of the minority class with respect to the overall size of the data for each project-version of the SDP dataset.
    
	\item \textbf{Class Overlap.}  We use Gong et al.'s~\cite{gong2019empirical} improved K-Means clustering
    cleaning approach (IKMCCA) to detect rows with overlapping classes for each project version of the SDP data. This approach solves both class overlap and class imbalance problems by combining the K-Means Clustering Cleaning Approach (KMCCA)~\cite{tang2004noise} and the Neighbourhood cleaning technique (NCL)~\cite{chen2018tackling}. Gong et al.~\cite{gong2019empirical} inculcated class imbalance into the KMCCA and empirically demonstrated model performance boosts from the IKMCCA over the KMCCA approach. After performing K-means on the explanatory variables, IKMCCA checks if the percentage of defective instances in the $i_{th}$ cluster lies below a threshold p\%, then the cluster's defective instances are removed, otherwise, the non-defective instances for that cluster are removed. \\
    Since class overlap depends on the labels, our studied dataset contains the potentially mislabeled variable (``\textit{HeuBug}'') and correctly labeled variable (``\textit{RealBug}''), we use the corrected labels in \textbf{RQ1}, and use the correct or mislabeled variables depending on the order of cleaning for building models in \textbf{RQ2} and \textbf{RQ3}.
 \end{itemize}
 
    \subsubsection{Feature-Row Imbalance}
    We check the ratio of the number of features (columns) and samples (rows) for each project version of our studied SDP dataset. The threshold of 10 is commonly used as a rule-of-thumb value in the industry\footnote{\url{https://graphite-note.com/how-much-data-is-needed-for-machine-learning}}\footnote{\url{ https://en.wikipedia.org/wiki/One_in_ten_rule}}.

    \subsubsection{Correlation \& Redundancy} 
    We consider both correlation \& redundancy as one group, since prior research~\cite{tantithamthavorn2016empirical} identified that these two have an analagous impact on model interpretation, while causing slight reduction to the model performance. While, typically, a data scientist/engineer would use domain knowledge to cherry pick one important feature from a group of the correlated and redundant ones, we use Jiarpakdee et al.'s~\cite{jiarpakdee2018autospearman} tool, , RNalytica~\cite{RNalytica}, specifically designed for picking the important SDP features from a cluster of correlated \& redundant variables.
    Moreover, this tool is commonly adopted by many research publications on software engineering~\cite{yatish2019mining,rajbahadur2021impact,tan2022does,hassan2018studying}. 

\subsection{Constructing Defect Models}
In this subsection, we discuss the different aspects of constructing ML models for \textbf{RQ2} and\textbf{ RQ3}.

\subsubsection{Studied Learners}
For \textbf{RQ2} and \textbf{RQ3}, we construct ML models using four learners, namely, \textit{Neural Network} (NN), \textit{Logistic Regression} (LR), \textit{Random Forest} (RF), and \textit{XG-Boost} (XGB). 
These learners are chosen as they represent the prominent families of learners~\cite{tantithamthavorn2016automated} used in defect prediction. Specifically, these learners were selected from the Bagging, Boosting, Regression, and Neural Network families respectively. 
From Tantithamthavorn et. al~\cite{tantithamthavorn2016automated}, we excluded other families such as Naive Bayes, Nearest Neighbors, Discrimination analysis, Rule-based, and SVM due to them being known for a lack of performance in defect prediction~\cite{yasir2022defect,khakhar2022integrity}. Since we already use the random forest classifier, which is an aggregation of decision trees, we exclude the decision tree family. Lastly, the Partial Least Square family is a dimensionality reduction technique, and is not directly applicable to our classification task. Within the Neural Network family, we use a plain neural network classifier with one to five layers, as opposed to RNN or CNN architectures given that deep learning was shown to not be necessarily better for the use case of SDP~\cite{zeng2021deep}.

\subsubsection{Hyper-parameter tuning}
Each model built for evaluating \textbf{RQ2} and \textbf{RQ3}, was subjected to hyper-parameter tuning for optimal model performance. We performed parameter tuning, keeping the $\#iterations=10$ (i.e., 10 different sets of hyper-parameters were evaluated) using the \textit{Random Searching} algorithm, which is more efficient than its alternative, \textit{grid search} in estimating the optimal hyper-parameters~\cite{bergstra2012random}. We kept the value of $k=3$ in k-fold cross-validation parameter (i.e., training data is split into three smaller folds and the process is repeated three times). Overall, the evaluation of one model's hyper-parameter with  $\#iterations=10$ and $k=3$ resulted in 30 model training runs. 

While software analytics research (e.g., Agarwal et al.~\cite{agrawal2019dodge}) has commonly used the default values of $k$, i.e., $k=5$, the massive number of models trained in our research, i.e., a total of 0.78 Million models for RQ2 and RQ3, made this value of $k$ infeasible, forcing us to reduce to $k=3$.

We obtain our hyper-parameter grid (see Table~\ref{tab.hyper-parameters}) by adopting values from popular software analytics research. Specifically, the hyper-parameters for LR and RF learners were adopted from Aggarwal et al.~\cite{agrawal2021simpler}, XGB hyper-parameters from Alazba et al.~\cite{alazba2022software}, and NN parameters from Bhaweres et al.~\cite{bahaweres2020software}. 

\begin{table}[]
\centering\caption{hyper-parameters for all learners used in this study.}\label{tab.hyper-parameters}
\begin{tabular}{|c|l|}
\hline
\textbf{learners} & \multicolumn{1}{c|}{\textbf{Hyper parameters}}                                                           \\ \hline
Random Forest &
  \begin{tabular}[c]{@{}l@{}}Criterion:   {[}gini, entropy{]}\\ N\_estimators:   randint(50, 150, 250, 500, 750)\\ Min\_samples\_split: randuniform(0,0.1)\end{tabular} \\ \hline
Logistic Regression & \begin{tabular}[c]{@{}l@{}}Penalty : {[}l2, none{]}, \\ C : random int (1, 500)\end{tabular}             \\ \hline
Neural Network &
  \begin{tabular}[c]{@{}l@{}}Learning\_rate: randuniform(0.001, 0.01)\\ Dropout: randuniform(0.1,0.5)\\ Neurons: randint(25,100)\\ Hidden\_layers: randint(1,5)\\ Batch\_size: {[}64{]}, Epoch: {[}15{]}\end{tabular} \\ \hline
XGB                 & \begin{tabular}[c]{@{}l@{}}N\_estimators: {[}100,50,500,1000{]}\\ Max\_depth: {[}6,3,4,5{]}\end{tabular} \\ \hline
\end{tabular}
\end{table}

\subsubsection{Model Validation}

We split 80\% of the data into training and 20\% into testing using stratified sampling to ensure that there are instances of buggy and non-buggy classes in both sets. This split is done 10 times, with ML models built on each split to obtain the performance metrics. Although bootstrapping is considered a robust validation strategy~\cite{rajbahadur2021impact}, sampling with replacement leads to duplication of data in the training set. This does not allow us to assess the impact of \textit{Duplication} as a data quality antipattern, restricting our use of bootstrapping as a validation technique.

\subsubsection{Performance Estimation}
 For each of the ML models built in \textbf{RQ2} and \textbf{RQ3}, we evaluate the performance of the SDP models using the two threshold dependent and four threshold independent metrics given below. A threshold classifies a probabilistic outcome as either ``defective'' or ``non-defective''.

\begin{itemize}
\item \textbf{AU-ROC} or \textit{Area Under the Receiver Operator Curve} is a threshold-independent metric. 
 AU-ROC measures the area under the curve at all thresholds of the receiver operator curve. This curve plots the true positive rate (sensitivity) against the false positive rate (\texttt{1 - Specificity}).

\item \textbf{AU-PRC} is another threshold-independent metric that calculates the \textit{Area Under the Precision-Recall Curve} at different thresholds. AU-PRC is a better estimate for datasets suffering from the class imbalance antipattern~\cite{davis2006relationship}.

\item \textbf{Pr} or \textit{Precision} estimates the observations correctly predicted out of all predictions. It is calculated as \texttt{True Positives / (True Positives + False Positives)}.

\item \textbf{Re} or \textit{Recall} estimates the observations correctly predicted over the total number of correct observations. It is calculated as \texttt{True Positives / (True Positives + False Negatives)}.

\item \textbf{\textit{F1 Score}} strikes a balance between Precision and Recall by calculating the harmonic mean between the two. Its calculated as \texttt{ 2(Pr * Re)/(Pr + Re)}

\item \textbf{MCC} or \textit{Mathews Correlation Coefficient} (MCC) is considered as a reliable metric~\cite{chicco2020advantages} as it produces good scores only if the prediction returns good rates for true positives (TP), true negatives (TN), false positives (FP), and false negatives (FN). MCC is calculated as\\ \texttt{$(TP*TN – FP*FN) / \sqrt{(TP+FP)(TP+FN)(TN+FP)(TN+FN)}$}\\

\end{itemize}
We use a threshold of 0.5 in the confusion matrix to estimate our threshold dependent metrics (Pr, Re, F1, and MCC). Although prior research suggests that threshold independent metrics (AU-ROC and AU-PRC) are more accurate estimators of model performance~\cite{tantithamthavorn2018experience,powers2020evaluation}, we report both to provide practitioners with more insights on how data quality antipatterns affect their specific use case. For example, in some cases, identification of True-Positives using Recall may be more important than AU-ROC or Precision.

%% file: rq1.tex
\subsection{\rqone} 
\noindent{\bf \textit{Motivation.}}

Despite the widespread use of SDP data in software engineering research~\cite{jiarpakdee2020empirical,jiarpakdee2019impact,tantithamthavorn2021explainable}, researchers utilizing SDP datasets lack an empirical understanding of the presence of data quality antipatterns within the software defect prediction datasets.
Singular data quality antipatterns have been found in  Software Engineering datasets (\cite{shepperd2013data,gong2019empirical,hynes2017data,wu2021data}), however, to the best of our knowledge, no research yet checked the prevalence of multiple data quality antipatterns in the context of software engineering. Understanding the prevalence of these antipatterns is important for both researchers and practitioners, as it can provide valuable insights into the factors that impact the performance and interpretation of machine learning models used for defect prediction. Hence, in this RQ, we aim to check the prevalence of data quality antipatterns in SDP data.\\

\noindent{\bf \textit{Results.}} Out of the 19, i.e., five top-level types\footnote{The top-level antipatterns having sub-types are not counted to avoid duplication in our counts} and 14 sub-types of antipatterns in our taxonomy~\ref{fig.antipattern.taxonomy}, we found nine antipatterns to commonly occur in the studied SDP dataset.
The distribution based antipatterns (tailed and unnormalized metrics) along with the label antipatterns, class overlap and mislabel are prevalent in all SDP project versions, while the other antipatterns occur in 2-50\% of the projects versions. 

We did not detect the presence of four antipatterns (i.e., \textit{Data Miscoding, Inconsistent Representation, Missing Values} and \textit{Data Drift}) in the studied datasets. 
Since \textit{Data Miscoding} is a dynamic issue, an ML pipeline (or non-ML data-driven program) should be able to validate the correct data type; nevertheless, we are unable to identify this as an antipattern in the static dataset from Yatish et al~\cite{yatish2019mining}. Furthermore, there are neither missing values nor inconsistent representations in the studied SDP data. Perhaps, Yatish et al.~\cite{yatish2019mining} cleaned missing values during the data procurement phase of conducting their study, while we do not suspect Yatish et al's work to encounter inconsistent representation, as this antipattern would most likely be prevalent in text-like features. 

Below, we present each of the antipatterns prevalent in the SDP data:

\subsubsection{Row Feature Imbalance}
\begin{figure}[!t]
          \centering
          \makebox[\textwidth][c]{\includegraphics[width=1.2\columnwidth, keepaspectratio]{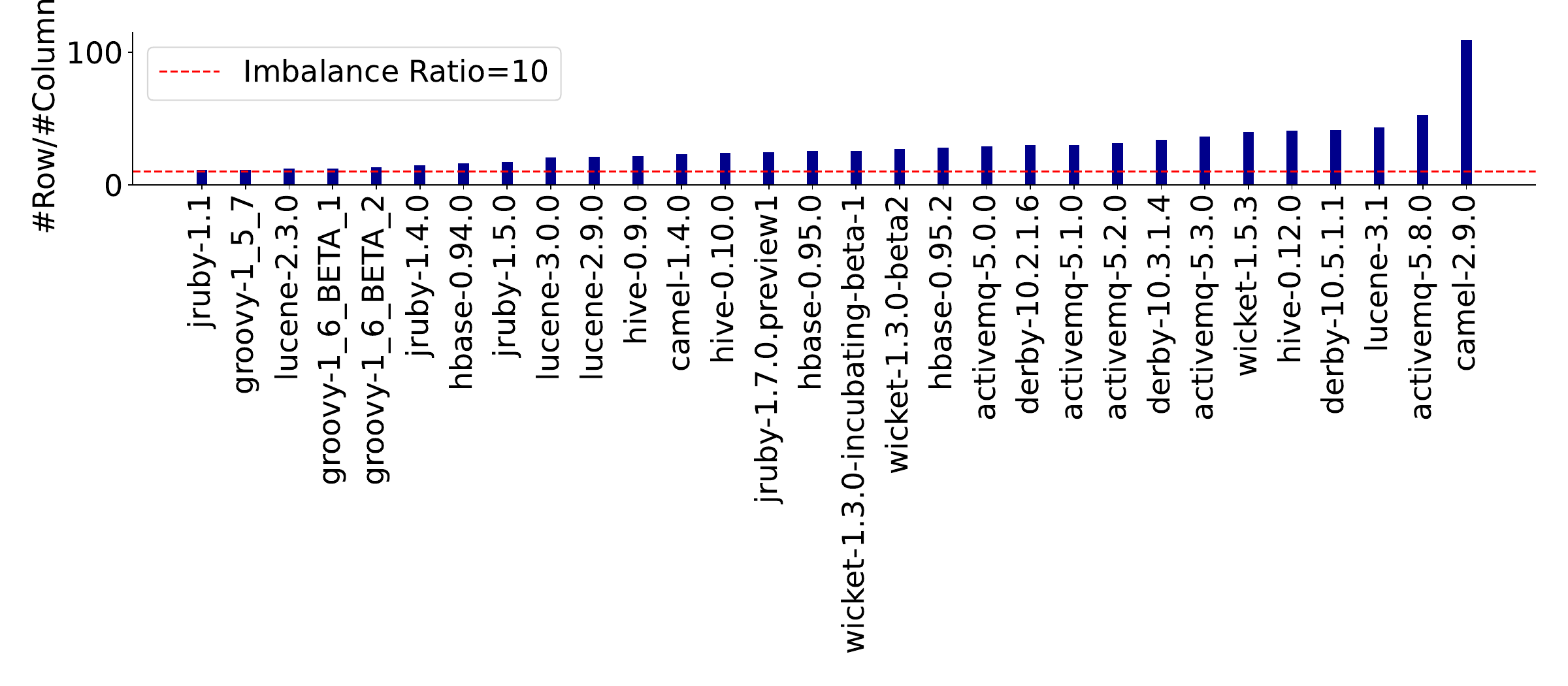}}
          \caption{Row feature imbalance. While each project version has 65 columns, the row size differs.}
          \label{fig.rq1.row_feat_imbalance}
        \end{figure}

\textbf{The studied SDP data lacks any row feature imbalance.} As illustrated in Figure~\ref{fig.rq1.row_feat_imbalance}, the ratio of the number of rows to the number of columns is greater than a threshold of 10, as demarcated by the red line in the Figure. The threshold of 10 is commonly used as a rule-of-thumb value in the industry\footnote{\url{https://graphite-note.com/how-much-data-is-needed-for-machine-learning}}\footnote{\url{ https://en.wikipedia.org/wiki/One_in_ten_rule}}. With more than 10 times the rows than the features, we consider there to be no row feature imbalance.
Notably, the outliers in Figure~\ref{fig.rq1.row_feat_imbalance} include  the project version \textit{Camel-2.9.0} having the maximum (7,120) number of rows, and the project version Jruby-1.1 having the minimum (731) number of rows. Each row in Yatish et al.`s~\cite{yatish2019mining} dataset represents a \texttt{.java} file, and these outliers are due to the project versions with the least or the maximum number of files in their codebase. 

\subsubsection{Schema violations} 
\begin{figure}[!th]
      \centering
      \includegraphics[width=0.5\columnwidth, keepaspectratio]{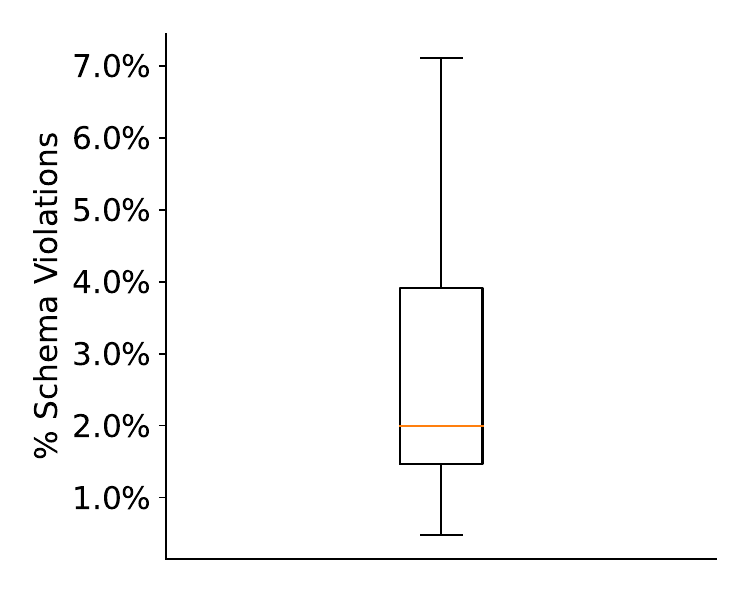}
      \caption{Distribution of \% Schema violations amongst the different studied project versions of the SDP data.}
      \label{fig.rq1.schema}
    \end{figure}

    \textbf{2\% of the studied SDP dataset has schema violations.} Figure~\ref{fig.rq1.schema} shows the distribution of violations for different project versions of the studied SDP data. We observed violations for five (i.e., \texttt{R8, R14, R15, R17, and R19} out of 20 rules defined in Table~\ref{tab.schema}. A total of 30, 12, 7, 6, and 6 files had violations against rules \texttt{R14, R8, R14, R15,} and \texttt{R17}. Overall, the maximum number of violations were observed for project \texttt{hive} version \texttt{0.10.0} with 7\% data having schema violations.

    We suspect that the calculation of complex metrics  like complexity could lead to higher chances of violations either due to errors in calculations or due to more complex code structures not adhering to the rules \texttt{R14, R15, R17}. Rules without violations might be simpler and more direct in their measurements (like \texttt{R1, R2, R3}). For \texttt{R14}, in deeply nested functions, calculation of the average number of lines might become skewed, with large disparity in size and complexity of nested functions which may have led to an erroneous calculation. Further work may be required to pinpoint the exact cause of such violations. 
    
\subsubsection{Data Distribution Anti-patterns}

\begin{figure}[!t] 
    \makebox[\textwidth][c]{\includegraphics[width=1.4\columnwidth, keepaspectratio]{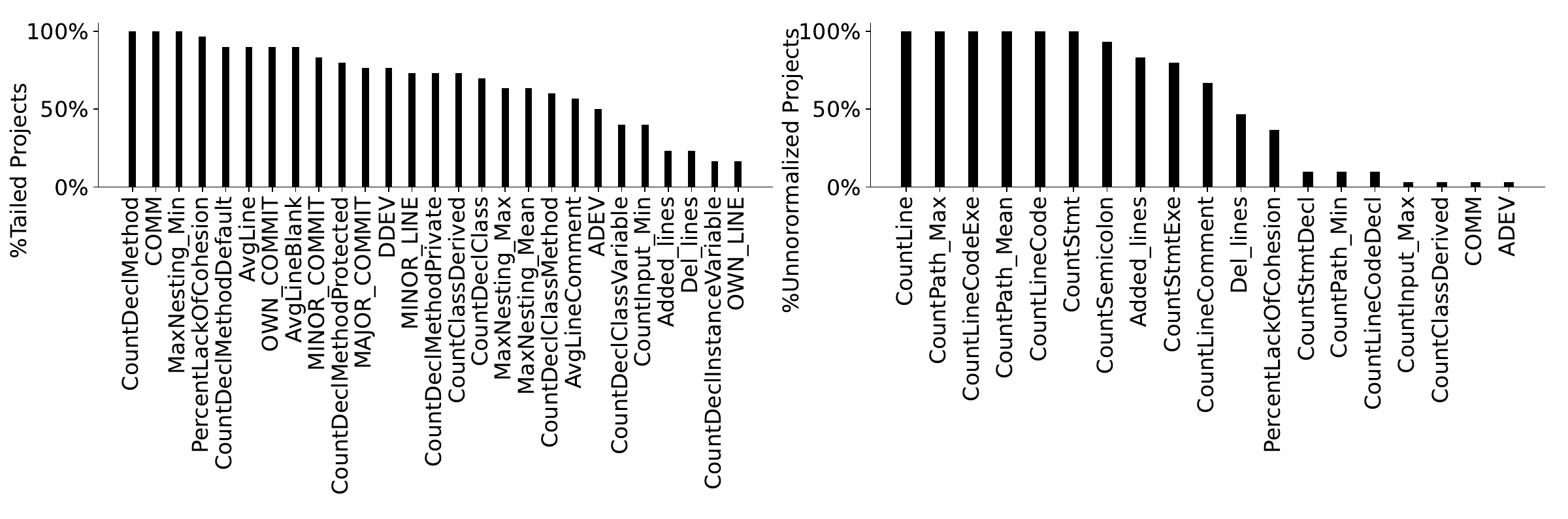}}
      \caption{Tailed metrics (left) along with Unnormalized metrics (right) across different project versions of the SDP data.}
      \label{fig.rq1.both_tailed_unnormalized}
\end{figure}    


\begin{itemize}
    \item \textit{Tailed Distribution.} \textbf{All project versions have tailed columns.} 
    The number of projects having tailed distributions is shown in Figure~\ref{fig.rq1.both_tailed_unnormalized} (left). Overall, the 5 point summary of tailed projects is : 11 (min), 15.25 (25th), 18 (median), 21 (27th 24 (max) indicating that the studied project versions have at least a median of 18 tailed metrics.    
    The metrics \textit{CountDeclMethod, COMM }and \textit{MaxNesting\_Min} are tailed across all projects. 

    \item \textit{Unnormalized Distributions.}\textbf{ Unnormalized distributions are prevalent in all project versions of the studied SDP dataset.}
    Particularly, 17\% of metrics are consistently unnormalized, while 37\% are inconsistently unnormalized. The 19 metrics detected as unnormalized are illustrated in Figure~\ref{fig.rq1.both_tailed_unnormalized} (right).

    \item \textit{Constant Features.} \textbf{We found that the \textit{Minor\_Commit} metric remained constant across five project-versions}: \textit{ActiveMQ-5.1.0, 5.2.0, 5.3.0, 5.8.0} and \textit{Derby-10.5.1.1}.

    \item \textit{Data Drift.} \textbf{We found no evidence of data drift between successive versions of the SDP data. }Furthermore, we found no drift between the first and final versions of the same project.

\end{itemize}

\subsubsection{Packaging Errors: Duplicated Data}
\begin{figure}[!th]
      \centering
      \includegraphics[width=0.5\columnwidth, keepaspectratio]{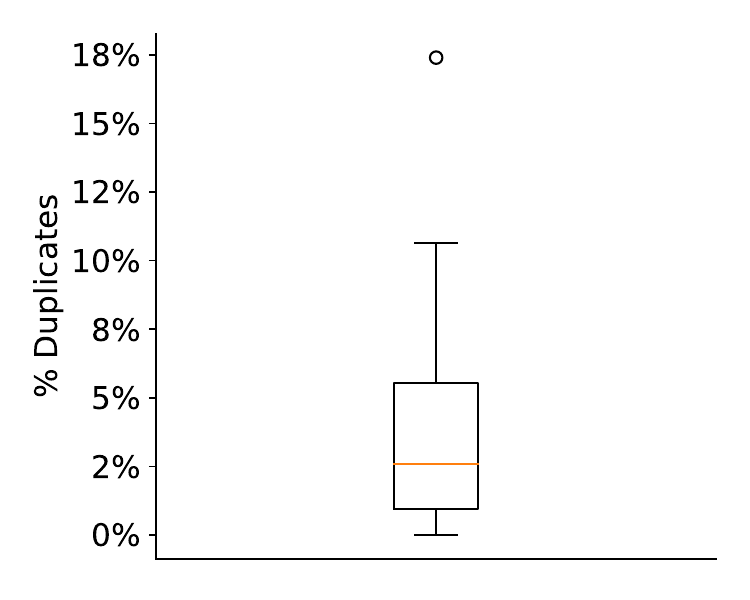}
      \caption{Percentage of duplicated rows detected in the studied SDP dataset. }
      \label{fig.rq1.duplicates}
    \end{figure}

    \textbf{In the studied SDP dataset, a median of 2.6\% data is duplicated.} Figure~\ref{fig.rq1.duplicates} depicts the percentage of duplicated rows across different project versions of the studied SDP data. The project \textit{ActiveMQ-5.8.0} contains the maximum, i.e., 17.4\% duplicates, whereas the project \textit{Hbase-0.94.0} contains no duplicates. 

\subsubsection{Label Antipatterns}
\begin{figure}[!th]
\makebox[\textwidth][c]{\includegraphics[width=1.2\columnwidth, keepaspectratio]{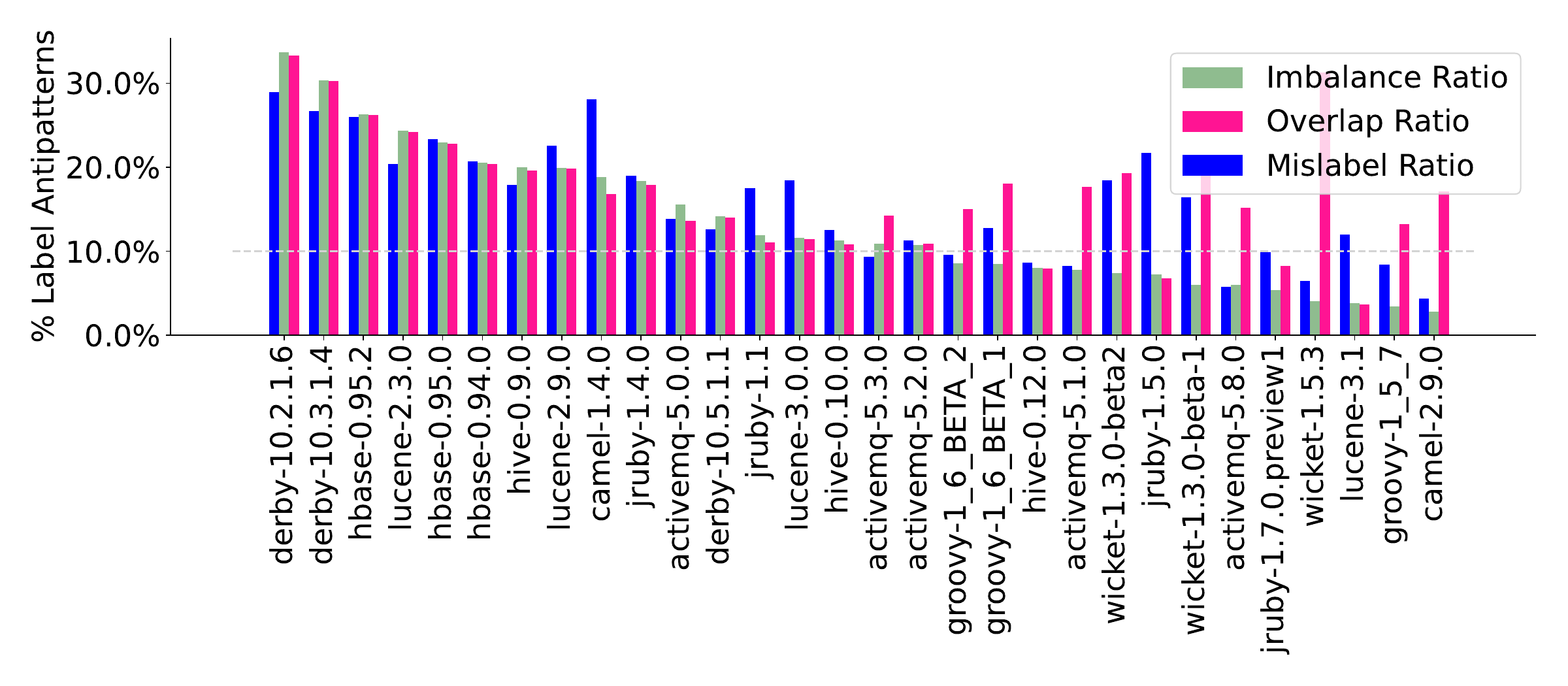}}
\caption{Distribution of Label Antipatterns in the SDP data. We present the ratio of class imbalance, class overlap along with mislabels.}
\label{fig.rq1.labelingErrs}
\end{figure}

\paragraph{Class Overlap.}
\textbf{A median SDP project version has 17\% rows having overlapping classes.} The percentage of overlapping instances detected on a per-project version basis is shown in red in Figure~\ref{fig.rq1.labelingErrs}. The project Derby, version 10.2.2.6, had the highest amount of overlapped instances (i.e., 654 out of 1,963). According to prior studies, such outcomes are to be expected in defect prediction~\cite{gong2019empirical,chen2018tackling}.

\paragraph{Class Imbalance.}
\textbf{A median of 11\% class imbalance is prevalent in the studied SDP dataset.} Figure~\ref{fig.rq1.labelingErrs} shows the ratio of minority classes (buggy instances) with respect to the total data size per project-version-basis. Class Imbalance is depicted by the blue color in Figure~\ref{fig.rq1.labelingErrs}. As shown in the Figure, project-versions having high class imbalance are also likely to have class overlap. Such findings are not uncommon as prior research has indicated the relation between the two antipatterns~\cite{johnson2019survey,yang2020rethinking,ali2019imbalance}. Below the threshold of 10\% as demarcated by the gray line in Figure~\ref{fig.rq1.labelingErrs}, 43\% of the project-versions have class imbalance. While a specific threshold for ``high'' class imbalance is not provided in literature, we use a threshold of 10\% to indicate the prevalence of class imbalance in different project-versions of the studied SDP dataset.
    
\paragraph{Mislabels.}
     \textbf{All project versions of the studied SDP dataset have mislabeling, as demonstrated by Yatish et al.~\cite{yatish2019mining}.}
     The inconsistencies observed between the two labeling schemes are depicted in green color in Figure~\ref{fig.rq1.labelingErrs}. A median of 15\% of labels had conflicting labels produced from the two (i.e., \texttt{Heuristic} or \texttt{Realistic}) labeling strategies.
     
\subsubsection{Correlation \& Redundancy}
\begin{figure}[!ht]
  \centering
  \includegraphics[width=0.5\columnwidth, keepaspectratio]{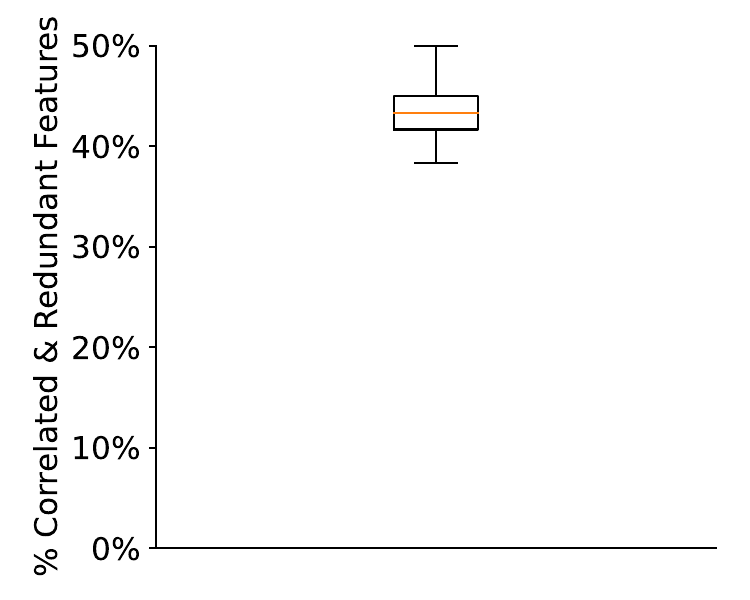}
  \caption{Correlated and redundant features in the studied SDP data.}
  \label{fig.rq1.corrredun}
\end{figure}
\textbf{Within the studied dataset, 38-50\% columns are correlated \& redundant.} Figure~\ref{fig.rq1.corrredun} shows the distribution of the percentage of correlated \& redundant features within different project versions of the studied SDP dataset. A median of 43.3\% metrics are correlated \& redundant. This is accompanied by a high variance of 8\%, which indicates that the quantity of correlated and redundant metrics varies across different project versions within the studied SDP dataset, indicating that each project has a unique distribution of such metrics. 


\begin{Summary}{Antipattern Prevalence}{}
Anti-patterns in the SDP data include \textit{
schema violations, correlated \& redundant features, distribution antipatterns (tailed, constant, and unnormalized distributions), labeling antipatterns (mislabeling, class overlap, class imbalance} and \textit{duplicated data}. We found no instances of \textit{data miscoding, missing values, row-feature imbalance} or \textit{data drift}.
\end{Summary}

\subsection{Overlap of Antipatterns}
\begin{figure}[!h]
\centering
\includegraphics[width=0.7\columnwidth, keepaspectratio]{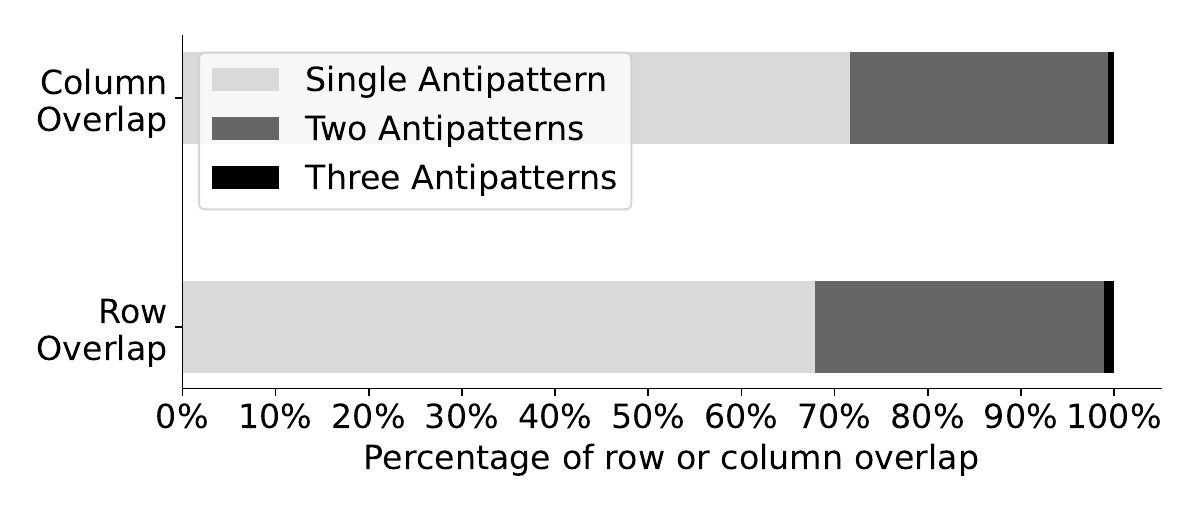}
\caption{Overlap between multiple antipatterns. We do not show the data points that do not have any antipattern.}
\label{fig.antipattern.overlap}
\end{figure}

In the previous subsection, we studied the prevalence of different antipatterns in our studied SDP dataset. In this subsection, we present the extent of overlap of these antipatterns. We bifurcate the results into two categories: row-level antipatterns (including \texttt{schema violations, mislabels, class overlapped rows,} and \texttt{duplicates}) and column-level antipatterns (encompassing \texttt{tailed, unnormalized, constant,} and \texttt{correlated \& redundant} features). We focus on the overlap within the same category, i.e., either row-level or column-level.

\begin{itemize}
    \item \textbf{Row level overlap. The majority (67.9\%) of rows impacted by  antipatterns do not exhibit any overlap.} As shown in Figure~\ref{fig.antipattern.overlap} only \textbf{31\% rows having antipatterns have two antipatterns, out of which 94\% cases are overlapped between Mislabels and Class Overlap}. The remaining 1\% cases having row-based antipatterns had three antipatterns affecting the same row. 

    \item \textbf{Column level overlap. Most (71.6\%) columns having antipatterns do not have any overlap with another antipattern} as shown in Fig~\ref{fig.antipattern.overlap}. 90\% of the remaining 28\% cases with two column-based antipatterns comprised correlation \& redundancy along with tailed features. A trivial amount of columns (0.4\%) had three antipatterns. 
\end{itemize}

\begin{Summary}{Antipattern Overlap}
For antipatterns having overlap on a row-level, 94\% of the cases comprise mislabels and class overlap. For the column-level antipattern overlap, 90\% of the cases include correlated \& redundant features along with tailed features.
    
\end{Summary}

%% file: rq2_orders.tex

\subsection{\rqtwoone}
\begin{figure}[!h]
        \makebox[\textwidth][c]{\includegraphics[width=\columnwidth, keepaspectratio]{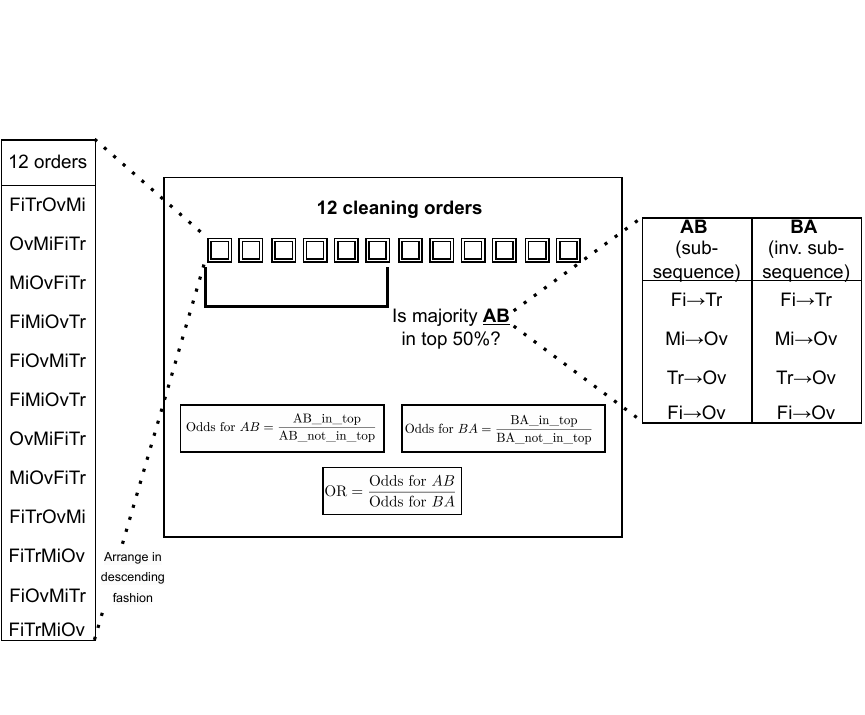}}
        \caption{Experiment setup for checking the impact of cleaning orders for various antipatterns. ``AB'' refers to cleaning A proceeded by cleaning B, while ``BA'' refers to the reverse order.}
        \label{fig.cleaningorder}
    \end{figure}
\noindent \textbf{\textit{Motivation.}} 
The data pre-processing phase in ML pipelines involves preparing the data for training the ML model, as noted by Amershi et al.~\cite{amershi2019software}. This includes cleaning, performing transformations, and more, a topic that has been widely studied in literature and is outlined in our Section~\ref{sec.back}. 
However, despite the recognition of the importance of cleaning data quality antipatterns, the optimal order for addressing multiple antipatterns remains unknown. Understanding the influence of the order in which antipatterns are cleaned is crucial, as it can affect the size of the training data and, thereby, the performance of the ML model. As demonstrated in our previous research question, 32.1\% (100-67.9)  rows and 28.4\% (100-71.6) columns with antipatterns overlap, indicating that data quality antipatterns can often coexist in practice, exacerbating this need. 

The overlap of these antipatterns may potentially lead to compounded errors and biases in the dataset, and  addressing them in an improper sequence may inadvertently amplify these issues, or negate the benefits of cleaning performed in previous steps. For instance, normalizing data before addressing missing values might lead to skewed distributions, impacting the accuracy of imputation techniques. Conversely, handling class overlapped rows prior to normalization could remove valuable data points that are actually significant once the data is scaled appropriately. Therefore, establishing an effective order for cleaning operations is essential to ensure that each step enhances the quality of the data without introducing new issues or magnifying existing ones.

\begin{table}[H]
\centering
\begin{tabular}{lcccc}
\hline
\textbf{Antipattern} & \textbf{Fi} & \textbf{Tr} & \textbf{Ov} & \textbf{Mi} \\ \hline
Schema Violations & \checkmark &  &  &  \\
Duplicated Data & \checkmark &  &  &  \\
Constant Features & \checkmark &  &  &  \\
Correlated \& Redundant & \checkmark &  &  &  \\
Log-Transformation &  & \checkmark &  &  \\
Z-Scoring &  & \checkmark &  &  \\
Mislabel Correction &  &  &  & \checkmark \\
Overlapping Classes &  &  & \checkmark &  \\ \hline
\end{tabular}
\caption{Categorization of Antipatterns under Cleaning Categories}
\label{tab:cleaning_categories}
\end{table}

\noindent \textbf{\textit{Approach.}} We remove antipatters in different orders from the training data and build ML models using the studied learners. Figure~\ref{fig.cleaningorder} provides an overview of our approach. A systematic description is provided below.

\begin{itemize}
    \item \textbf{Cleaning Types.}
    We categorize the cleaning into four cleaning types: \textit{Filtering} (Fi), \textit{Transformations} (Tr), \textit{Class Overlap} (Ov), and \textit{Mislabeling} (Mi). We use the abbreviations for each type henceforth. Classification into these types is based on the type of cleaning required for each antipattern as shown in Table \ref{tab:cleaning_categories}. 
    For instance, the \textit{Fi} category involves filtering schema violations, duplicated data, constant features, and correlated and redundant features, as data is removed (filtered) from rows or columns. The \textit{Tr} category entails performing log-transformation and z-scoring to transform tailed and unnormalized features respectively. 
    We keep \textit{Class Overlap} separate from other filtering-based cleanings because \textit{Mislabeling} explicitly impacts \textit{Class Overlap}, while having no effect on other cleaning techniques, such as schema violations or constant features.

    \item \textbf{Creating cleaning orders.}
    Performing different order of cleaning entails performing one type of cleaning first, and other types later. For example, an order \textit{FiTrMiOv} would entail performing cleaning in the following order Fi→Tr→Mi→Ov. Similarly, from the four cleaning groups \textit{Fi, Tr, Mi} and \textit{Ov} lead to 4$!$, i.e., 24 cleaning permutations (e.g., \textit{FiTrMiOv, FiTrOvMi, TrFiMiOv, etc.}). We call these permutations as ``orders'' henceforth.
    
    \item \textbf{Culling redundant orders.} From all the permutations, we exclude those that lead to the same order. In particular, the positioning of \textit{Mi} at any step before \textit{Ov} does not matter, as both \textit{Mi} and \textit{Ov} are based on the response (y-label) and do not impact the explanatory variables. For instance, in the case of \textit{FiTrMiOv}, mislabel positioning before overlap (i.e., \textit{FiMiTrOv}, or \textit{MiFiTrOv}) yields the same data. Similarly, mislabel correction after class overlapping (i.e., \textit{OvFiTrMi, OvFiMiTr, OvMiFiTr}) yields the same data as well. Removing such redundant orders leaves us with 12 orders. \emph{These cleaning orders are: \textit{FiMiOvTr, FiOvMiTr,  FiTrMiOv, FiTrOvMi, MiOvFiTr, OvMiFiTr, MiOvTrFi, OvMiTrFi, TrFiMiOv, TrFiOvMi, TrMiOvFi}, and \textit{TrOvMiFi}.} 
    
    \item \textbf{Building ML models from each cleaning order.} We clean the training data based on each of these orders, and build an ML model using the model-building process outlined in Section~\ref{sec.casestudy}. Using the unseen test set, we obtain model metrics, Precision, Recall, F1, MCC, AU-ROC, and AU-PRC from each model. To enable fairness, we compare the performance of models built using different cleaning orders of the same training data and evaluate the performance on the same unseen test set. 

    \item \textbf{Creation of sub-sequences.}
    Our objective is to evaluate whether one cleaning type outperforms another cleaning type.
    To study this, we repeatedly divide the 12 orders into two groups (sub-sequences) based on the occurrence of an antipattern pair - one group containing all orders in which an antipattern precedes another, and the other group with all orders in which the antipattern follows another.
    For instance, to check whether \textit{Fi} should be performed before or after performing \textit{Tr}, we form the sub-sequence \textit{Fi→ Tr} along with its inverse sub-sequence  \textit{Tr → Fi}. The sub-sequence \textit{Fi→ Tr} has six orders in which Filtering is performed before Transformation (i.e., \textit{FiMiOvTr, FiOvMiTr,  FiTrMiOv, FiTrOvMi, MiOvFiTr, OvMiFiTr}), while its inverse sub-sequence, \textit{Tr → Fi} has the remaining six orders. Similarly, from the 12 orders, six belong to one sub-sequence and the remaining six belong to the inverse sub-sequence. In total, we have four sub-sequences, i.e., MiOv, FiTr, TrOv, and FiOv; and naturally their inverse sub-sequences would be OvMi, TrFi, OvTr, and OvFi respectively.

    \item \textbf{Evaluation of sub-sequence impact.}
    We use the odds ratio (OR) \cite{bland2000odds} to show the likelihood of one sequence leading to better performance over its inverse (e.g., Fi→Tr versus Tr→Fi). The null hypothesis is that a sequence $ab$ ($a → b$) is in the top-performing sequences for a given set of training data. The alternative hypothesis is that a sequence $ab$ is not in the top 50\% performing sequences for the same training data. Hence, we calculate the odds ratio as follows for all models in a learner:
    
        \[ \text{Odds for } AB = \frac{\text{AB\_in\_top}}{\text{AB\_not\_in\_top}} \]
   
    Similarly,
        \[ \text{Odds for } BA = \frac{\text{BA\_in\_top}}{\text{BA\_not\_in\_top}} \]

    To calculate the odds ratio, we use:
    
        \[ \text{OR} = \frac{\text{Odds for } AB}{\text{Odds for } BA} \]

    An odds ratio of 1 indicates that the odds of $AB$ and $BA$ are the same. To interpret the significance of the odds ratios, we set thresholds where an OR in the range of 0.64 to 1.5 is considered unimportant~\cite{tenny2017odds,grimes2008making}. This implies that variations within this range do not significantly affect the performance difference between the sequences. Conversely, an OR less than 0.64 or greater than 1.5 is deemed important, signifying a substantial difference in performance. Overall, the experiment setup is given in Figure~\ref{fig.cleaningorder}. 
    


\end{itemize}

\begin{figure}[!h]
      \makebox[\textwidth][c]{\includegraphics[width=\columnwidth, keepaspectratio]{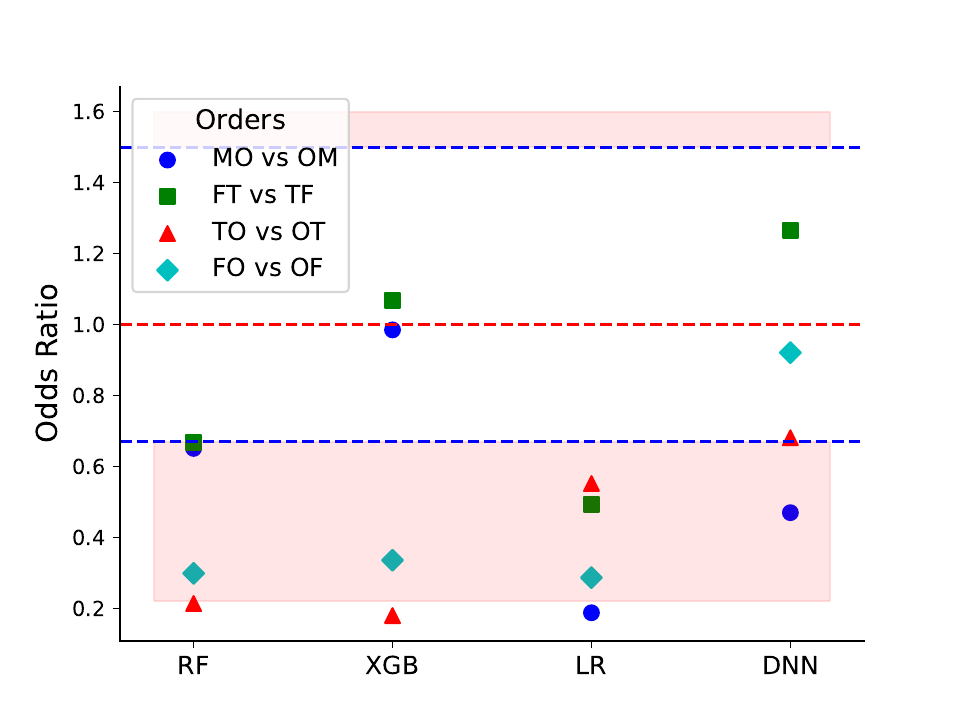}}
      \caption{The odds ratio of a given order with respect to its inverse order. The shaded area (demarcated by the red area) at 1.5 indicates a significantly high odds ratio, while a value below 0.66 indicates a significantly low odds ratio.
      }\label{fig.orders_odds_results}
    \end{figure}

\noindent \textbf{\textit{Results.} Different learners (RF, XGB, LR, DNN) respond uniquely to the order of cleaning antipatterns.} Figure~\ref{fig.orders_odds_results} illustrates the odds ratios for different orders of cleaning antipatterns. Notably, the odds ratios for orders Mi→Ov, Fi→Tr, and Tr→Ov vary across learners, highlighting the learner-dependent impact of cleaning orders. DNN, having higher degrees of freedom is the least impacted by the cleaning order. Below, we explain the rationale for this variation for each pair of cleaning orders, along with the differences in the learners below.

\subsubsection{MiOv vs OvMi}
Correcting mislabeled instances in the dataset primarily improves the quality of the target variable (y-label). If mislabels are not corrected first, overlap removal might incorrectly remove or retain instances based on inaccurate labels.

As indicated by the blue circles in Figure~\ref{fig.orders_odds_results}, \textbf{logistic Regression is highly sensitive to the quality of class labels.} Correcting mislabels first ensures that the subsequent removal of overlapping classes is based on accurate labels, which is crucial for LR as it directly models the probability of the target variable. This sequence improves model performance, resulting in a higher odds ratio for MiOv compared to OvMi. \noindent \textbf{Random Forest (RF) and XGBoost (XGB), on the contrary, are more robust to noise and label errors due to their aggregative nature}. Being ensemble based methods, RF and XGB can mitigate the impact of mislabels through averaging and boosting mechanisms, perhaps making the order of MiOv vs OvMi less critical. This robustness explains the odds ratios closer to 1, indicating less sensitivity to the sequence of these operations. Finally, \textbf{DNNs have a high degree of freedom and can model complex patterns}. They can adapt to initial mislabels or overlaps, making them less sensitive to the order of Mi and Ov. The results in Figure~\ref{fig.orders_odds_results} show that DNNs can handle different cleaning orders with minimal impact on performance, reflected by the odds ratios near 1.

\subsubsection{FiTr vs TrFi}
Since the Fi aims to remove extreme values or incorrect data points, performing Fi before Tr would help in reducing the impact of these outliers, which might still influence the transformed data if done post-transformation (TF). \\
As shown in the Green rectangles in Figure~\ref{fig.orders_odds_results}, \textbf{RF,  XGB and DNN are less sensitive to the order of Tr and Fi cleanings.} XGB is known for its robustness to different feature scales due to its grading boosting framework and can handle missing values and outliers more effectively making it less sensitive to the order of Fi and Tr. Similarly, DNNs are highly flexible for learning complex data patterns, given their higher degrees of freedom. Out of these three learners, RF is the least resilient to the order of Fi and Tr given the  its odds ratio at the border-line significance level in Figure~\ref{fig.orders_odds_results}. Perhaps, transforming later (FiTr) can remove the variability needed by the decision trees within the random forest, whereas, due to their robustness (given RF is an ensemble-based technique), the order has a less pronounced impact, with the odds ratios closer to 1.\\
In fact, filtering after transformation would 1) lead to incorrect filtering since the rule based checks would not work on transformed data; and 2) invalidate transformations (e.g., filtering data after log transformation would change the distribution of the logged data). However, Tr first improves the performance of LR and RF (slightly), as shown by the lower odds ratio for FT compared to TF. This can be attributed to the reduced influence of outliers. Post-transformation, the effects of outliers on the model’s learning process are reduced, because transformations like the logarithm reduce the variability caused by extreme values. Hence, even though these outliers are still present in the dataset during the training phase, their influence is diminished, allowing the models to focus on more general patterns. LR having the lowest DOF amongst the other learners makes the impact of the FiTr order the most pronounced.

\subsubsection{TrOv vs OvTr}
\textbf{Removing overlapped datapoints before transformation boosts the performance for all learners.} The odds ratios significantly less than 1 indicates that performing overlap removal first (OT) results in comparatively better performance than performing transformation first (TO). Class overlap removal entails removing a significant number of data points (as shown in RQ1), which, if done after the transformation, makes the transformations no longer meaningful. For instance, once a feature's log transformation is done, followed by the removal of 17\% data due to class overlap, the distribution of that feature is no longer a logged distribution. Conversely, from the perspective of class overlap removal on model performance, the data distribution available for class overlap for an accurate removal, should be similar to that of the final data used to train the model. When transformations are done later, the distribution based on which class overlap removal was done, and the distributions based on which a model is trained are dissimilar to each other, which could also explain the lower performance for OvTr as compared to TrOv.\\
Similar to the case of TrFi, the effects are less pronounced for DNN, given its ability to be less resilient to noise.

\subsubsection{FiOv vs OvFi}
RF, XGB, and LR have odds ratios significantly less than 1, suggesting that performing overlap removal before filtering tends to result in better performance compared to the reverse order, whereas DNN, with an odds ratio close to 1, indicate that the sequence of these operations does not significantly impact performance. We suspect \textbf{data dependency} and \textbf{model selectivity} to influence the performance of the Filtering and Overlap removal order.\\
\textit{Data Dependency}: The sequence might affect how effectively each method can leverage the structure of the data. For instance, overlap removal might expose patterns that filtering can utilize more efficiently afterwards.\\
\textit{Model Sensitivity}: Learners like LR and decision trees (used in RF and XGB), might be more sensitive to the presence of outliers or overlapping classes early in the cleaning process, whereas deep neural networks might be better at handling unstructured or noisy data due to their complexity and capacity for pattern extraction.

\subsubsection{Learner Complexities and Cleaning Order Impact}
In the previous subsections, we explored the impact of each order on the performance of our studied models. This section delves into why different learners exhibit varied responses to these operations, particularly focusing on their inherent characteristics and degrees of freedom. A degree of freedom (DOF) in ML is a factor indicating the number of parameters or features a model can adjust to fit the data, which in turn measures the model's complexity and flexibility.  In LR, for example, the number of degree of freedom is $n+1$, where $n$ is the number of features, accounting for each feature's coefficient and the intercept. This relatively lower DOF renders LR more sensitive to the order of data cleaning operations and particularly more prone to performance declines when the order does not align with optimal preprocessing for class label quality. 

RF and XGB, feature a more complex DOF dependent not just on the number of trees and their depth but also factors like learning rate and regularization (particularly for XGB), and exhibit a higher resilience to variations in data cleaning sequence. Their ensemble nature helps them manage label noise and class overlap effectively irrespective of the order. Conversely, DNNs have the highest DOF, with numerous weights and biases across multiple layers, enabling them to adapt flexibly to almost any data cleaning order. This capacity is reflected in their odds ratios near 1, as shown in Figure~\ref{fig.orders_odds_results}, highlighting their robustness to the sequence of cleaning operations. The distinct ways these models manage bias and variance, influenced significantly by their DOF, underpin their varied performance across different cleaning orders.

\begin{Summary}{Summary of RQ2}{}
The order of cleaning impacts model performance, with certain orders having more pronounced effects than others. Removing overlaps before  transformation (OvTr) consistently boosts the performance across all learners, emphasizing the importance of performing transformation at the end. In contrast, sequences like filtering before transformation (FiTr) and mislabel correction before overlap removal (MiOv) show varied impacts depending on the learner; LR benefits significantly from MiOv due to its reliance on label accuracy, while DNN shows robustness to the order of cleaning, highlighting its capacity to be resilient to different orders of cleaning antipatterns.
\end{Summary}

%% file: RQ2_aps.tex
\subsection{\rqtwotwo} 
\noindent \textbf{\textit{Motivation.}} 

While prior studies~\cite{gong2019empirical,shepperd2013data,yatish2019mining,mcintosh2016empirical} have presented the impact of the prevalence of an antipattern on model performance when other antipatterns are present, it is unclear what the impact of the prevalence of an antipattern is when the other antipatterns have been cleaned out. Understanding this relationship is important because in practical scenarios, multiple antipatterns are present, which may have overlapping effects. This raises the question whether fixing an antipattern is necessary when other antipatterns have been removed, given the ``cost'' associated with antipattern removal. For instance†, as evidenced by \textbf{RQ1}, a median of 17\% rows need to be removed to combat class overlap. 

Similarly, on the column level, removing correlated \& redundant features in \textbf{RQ1} led to a data loss of a median of 43\% features in our studied dataset. Indeed, noisy data can still contribute valuable information to ML models, particularly when clean data is scarce or unavailable~\cite{qi2018impacts,frenay2013classification,bootkrajang2012label}. Hence, to understand the value of removing a specific antipattern, it is important to assess the impact of a singular particular antipattern in a scenario where the remaining antipatterns are removed.

\begin{figure}[t]
        \centering
        \makebox[\textwidth][c]{\includegraphics[width=1.3\columnwidth, keepaspectratio]{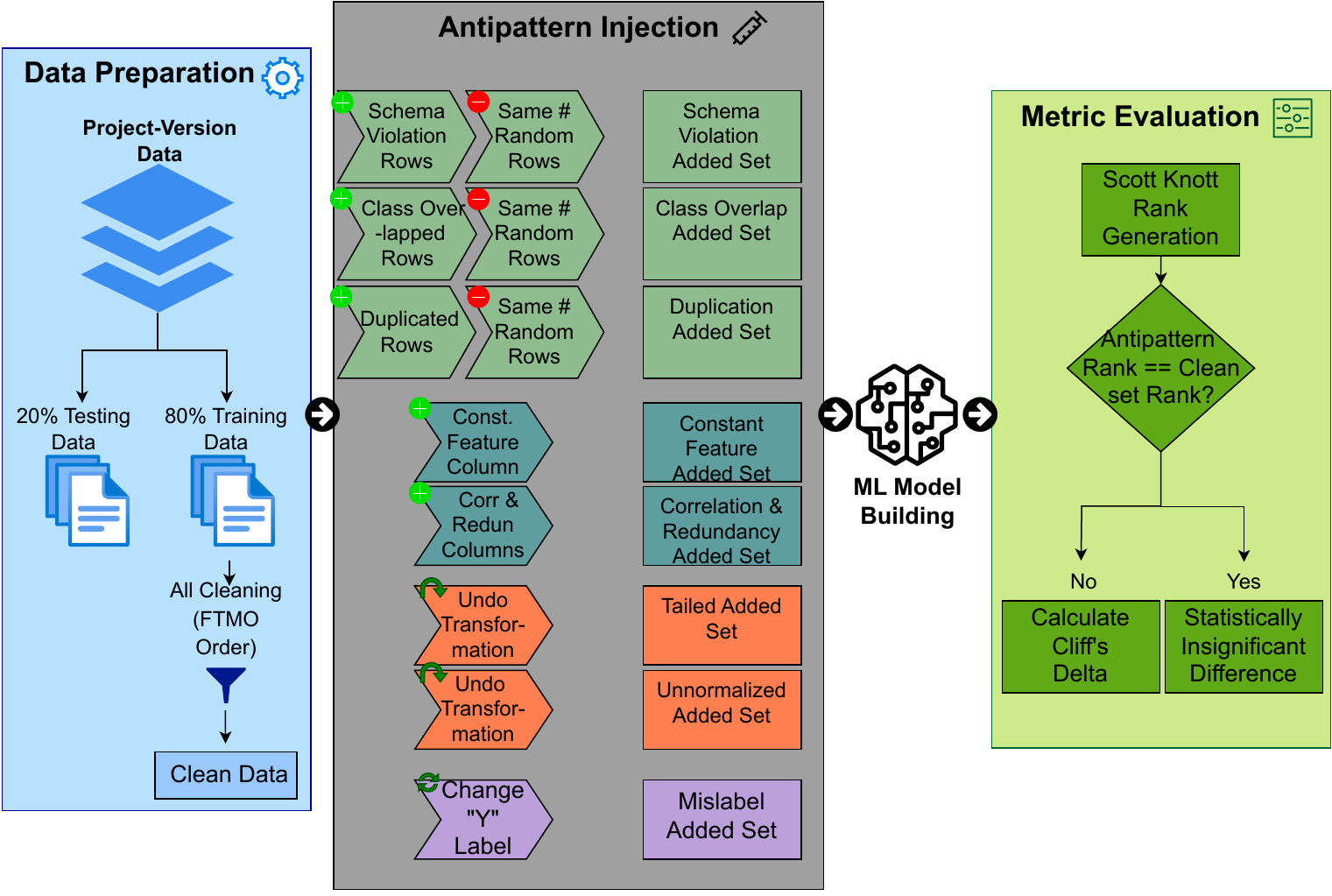}}
        \caption{Approach for assessing the impact of different data quality antipatterns. The ``removal of random rows'' is done to keep the data size consistent.}
        \label{fig.aps_approach}
        \end{figure}

\noindent \textbf{\textit{Approach.}}
Figure~\ref{fig.aps_approach} illustrates the high-level design of our approach. Overall, for each version of the project, we split the data into training (80\%) and testing (20\%) sets as mentioned in Section~\ref{sec.casestudy}. For each training data, we create a clean set by removing all antipatterns using the \textit{FTMO} order. Next, we add back one antipattern and build ML models and assess the model performance. A detailed explanation of our experiment setup is provided below.

\begin{enumerate}
\item \textit{Generation of a baseline ``clean'' set.} We generate an ``all-clean'' set which is devoid of any antipattern. This is shown as the output of the ``data-preparation'' step, shown in blue in Figure~\ref{fig.aps_approach}. From the various data cleaning orders presented \textbf{RQ2}, we chose the order \textit{FTMO} (i.e., $Filter \rightarrow Transform \rightarrow Mislabel \rightarrow Overlap$) based on the following rationale:

    \begin{itemize}
        \item In order to addressing schema violations, such as ensuring that column values meet specified thresholds, column dependencies, or minimum value requirements, it is crucial to conduct rule-based checks before performing transformations like z-scoring and logging. Hence, \textit{filtering should be performed before transformation.}
        \item Moreover, if a certain feature will be eliminated as a correlated or redundant feature, it doesn't make sense to transform that feature. As a result, \textit{filtering should be performed before transformation. }
        \item Class overlap should have mislabel correction applied to ensure that the class overlap detection has  accurate knowledge of the response label. Hence, \textit{class overlap should be performed after mislabel correction.}
        \item To ensure that the data distribution available for detection of overlapping classes is the same as that used for model training, \textit{class overlap should be carried out at the end.}
    \end{itemize}    

\item \textit{Adding back antipatterns.} 
To assess the impact of each antipattern, we add back one antipattern at a time into the clean data. This is shown in the gray ``antipattern injection'' part of Fig~\ref{fig.aps_approach}. In particular, we generate datasets with schema violations, duplicates, and class overlap by adding back these antipattern on the row-level (green color in Fig~\ref{fig.aps_approach}). On the column-level, we add new antipattern columns for correlated \& redundant features, and constant features (teal color in Fig~\ref{fig.aps_approach}). For the column-level antipatterns, tailed and unnormalized features, we do not perform the transformations (i.e., log for tailed and z-score for unnormalized) that were initially made as corrections to create the clean set (orange color in Fig~\ref{fig.aps_approach}). While injecting the mislabel antipattern, we train the model on the mislabeled data (purple color in Fig~\ref{fig.aps_approach}), while testing on the correct data. 

While adding row-based antipatterns into the clean set, the data size (i.e., row count) changes. The fact that model performance is dependent on the data size~\cite{domingos2012few,halevy2009unreasonable} can add bias to our results.
To eliminate this bias, we control for the data size by randomly removing clean data while adding back the row-based antipattern to maintain a consistent size of all antipattern-specific datasets. For example, while creating the schema violation dataset, we add back the $k$ schema violations into the clean set along with randomly removing $k$ clean rows to keep the size consistent as that of the clean set. 

\item \textit{Model generation and metric evaluation.}
We build machine learning models for four learners using the settings described in Section~\ref{sec.casestudy}. We generate F1, Precision, Recall, MCC, AU-ROC, and AU-PRC metrics on the unseen test set on the 10 folds of each project version data.  Since ML pipelines use the same preprocessing components for training and testing infrastructures~\cite{amershi2019software,bhatia2022towards}, we perform the same two transformations (i.e., z-scoring or log transformation) on the test set as that of the training set.\\
We use the Scott-Knott ESD~\cite{tantithamthavorn2016empirical} ranking algorithm to compare the metrics of the antipattern-containing sets to the clean set and determine if there is a statistically significant difference with respect to the clean set. The ESD is an effect-aware variant of the traditional Scott Knott test, which uses hierarchical clustering to partition different sets of data means into statistically distinct groups. We adopt the value of $\alpha=0.05$ from prior research~\cite{tantithamthavorn2016empirical}. 
For each project-version of the SDP data, we evaluate 10 trained models using various metrics. Each metric is evaluated on different performance groups, including \textit{tailed,  unnormalized, constant feature, duplicates, schema, correlation \& redundancy, mislabel, class overlap}, and a \textit{clean} set. If an antipattern has a different rank as compared to the clean set, we consider it to be statistically significant. For each significant comparison, we also calculate the effect size.

\item \textit{Effect size for significant comparisons.}
We use Cliff's $\delta$~\cite{cliff2014ordinal}as a measure of effect size to compare two groups of scores and determine their difference in magnitude. Ranging from -1 to 1, a strong effect size is indicated by values close to +/-1 (i.e.,+1 indicates that all values in one group are larger than the other, conversely, -1 indicates all values in one group are smaller) and no or negligible effect size by values close to 0. Cliff's $\delta$ is preferred over traditional measures such as Cohen's d, as it is more robust and insensitive to sample size, making it suitable for small sample sizes. We present the rank difference (R $\delta$) and Cliff's $\delta$ with respect to the clean set, with negative values indicating improvement from the clean set. We use the following interpretation of Cliff's $\delta$~\cite{cliff2014ordinal}. 

    \setlength{\abovedisplayskip}{0pt}
    \begin{align*}
    \text{Cliff's } \delta \text{ Effect Size} = 
        \begin{cases}
            \text{Negligible,} & \text{if } |\delta| \leq 0.147 \\
            \text{Small,} & \text{if }  0.147 < |\delta| \leq 0.33 \\
            \text{Medium,} & \text{if }  0.33 < |\delta| \leq 0.47 \\
            \text{Large} & \text{if }  |\delta| > 0.474
        \end{cases}
    \end{align*}
    \end{enumerate}

\begin{table}[!h]
\caption{Statistically significant metrics for all four learners. The abbreviation ``R $\delta$'' is used for rank difference w.r.t. the clean set, ``Effect'' for Cliff's $\delta$ Effect Size, and ``AP'' for the Antipattern that is added to the clean set. Large and medium effect sizes are bolded.
}\label{tab.ap}
\hspace{-2cm}\begin{tabular}{r lrl lrl lrl lrl}
\toprule
      & \multicolumn{3}{c}{\textbf{Neural Network}} & \multicolumn{3}{c}{\textbf{Logistic regression}} & \multicolumn{3}{c}{\textbf{Random Forest}} & \multicolumn{3}{l}{\textbf{Xg Boost}} \\
\midrule
\multicolumn{1}{l}{\textbf{Metric}} & \textbf{AP} & \multicolumn{1}{l}{\textbf{R $\delta$}} & \textbf{Effect} & \multicolumn{1}{l}{\textbf{AP}} & \multicolumn{1}{l}{\textbf{R $\delta$}} & \multicolumn{1}{l}{\textbf{Effect}} & \multicolumn{1}{l}{\textbf{AP}} & \multicolumn{1}{l}{\textbf{R $\delta$}} & \multicolumn{1}{l}{\textbf{Effect}} & \multicolumn{1}{l}{\textbf{AP}} & \multicolumn{1}{l}{\textbf{R $\delta$}} & \multicolumn{1}{l}{\textbf{Effect}} \\

\midrule

\multicolumn{1}{c}{\textbf{Precision}} & Tailed & 1     & S(-0.31) & \multicolumn{1}{l}{Tailed} & 2     & \multicolumn{1}{l}{\textbf{L (-0.48)}} & \multicolumn{1}{l}{Cls Ovr} & -1    & \multicolumn{1}{l}{S(0.17)} & \multicolumn{1}{l}{Cls Ovr} & -1    & \multicolumn{1}{l}{S(0.18)} \\

\midrule

\multicolumn{1}{c}{\multirow{4}[0]{*}{\textbf{Recall}}} & Tailed & 2     & \textbf{L(-0.52)} & \multicolumn{1}{l}{Unnrm} & 1     & \multicolumn{1}{l}{N(-0.06)} & \multicolumn{1}{l}{Cls Ovr} & 1     & \multicolumn{1}{l}{\textbf{M(-0.35)}} & \multicolumn{1}{l}{Tailed} & 1     & \multicolumn{1}{l}{N(-0.13)} \\
      & Mislabel & -1    & S(0.2) & \multicolumn{1}{l}{Mislabel} & -1    & \multicolumn{1}{l}{S(0.16)} &       &       &       & \multicolumn{1}{l}{Nt Nrm} & 1     & \multicolumn{1}{l}{N(-0.13)} \\
      & Cls Ovr & 1     & N(-0.11) & \multicolumn{1}{l}{Cls Ovr} & 1     & \multicolumn{1}{l}{N(-0.14)} &       &       &       & \multicolumn{1}{l}{Cor Rdn} & 1     & \multicolumn{1}{l}{N(-0.1)} \\
      & Cls Imb & 1     & S(-0.19) & \multicolumn{1}{l}{Cls Ovr} & 1     & \multicolumn{1}{l}{N(-0.14)} &       &       &       & \multicolumn{1}{l}{Cls Ovr} & 2     & \multicolumn{1}{l}{S(-0.33)} \\

\midrule

\multicolumn{1}{c}{\multirow{4}[0]{*}{\textbf{F1}}} & Tailed & 1     & S(-0.33) & \multicolumn{1}{l}{Tailed} & 1     & \multicolumn{1}{l}{\textbf{M(-0.33)}} &       &       &       & \multicolumn{1}{l}{Cor Rdn} & -1    & \multicolumn{1}{l}{N(0.1)} \\
      & Cls Imb & 1     & S(-0.18) & \multicolumn{1}{l}{Cor Rdn} & -1    & \multicolumn{1}{l}{N(0.07)} &       &       &       & \multicolumn{1}{l}{Mislabel} & -1    & \multicolumn{1}{l}{N(0.09)} \\
      &       &       &       & \multicolumn{1}{l}{Mislabel} & -1    & \multicolumn{1}{l}{N(0.1)} &       &       &       &       &       &  \\
      &       &       &       & \multicolumn{1}{l}{Cls Ovr} & -1    & \multicolumn{1}{l}{S(0.19)} &       &       &       &       &       &  \\

\midrule

\multicolumn{1}{c}{\multirow{3}[0]{*}{\textbf{MCC}}} & Tailed & 1     & \textbf{M(-0.39)} & \multicolumn{1}{l}{Tailed} & 1     & \multicolumn{1}{l}{\textbf{M(-0.43)}} & \multicolumn{1}{l}{Cls Ovr} & -1    & \multicolumn{1}{l}{S(0.2)} & \multicolumn{1}{l}{Cor Rdn} & -1    & \multicolumn{1}{l}{S(0.16)} \\
      & Cor Rdn & -1    & S(0.2) & \multicolumn{1}{l}{Cls Ovr} & -1    & \multicolumn{1}{l}{S(0.31)} &       &       &       & \multicolumn{1}{l}{Cls Ovr} & -2    & \multicolumn{1}{l}{S(0.3)} \\
      & Cls Ovr & -1    & S(0.22) &       &       &       &       &       &       &       &       &  \\

\midrule

\multicolumn{1}{c}{\multirow{4}[0]{*}{\textbf{AU-PRC}}} & Tailed & 1     & S(-0.16) & \multicolumn{1}{l}{Nt Nrm} & 1     & \multicolumn{1}{l}{N(-0.12)} & \multicolumn{1}{l}{Tailed} & 1     & \multicolumn{1}{l}{N(-0.14)} &       &       &  \\
      & Cor Rdn & -1    & S(0.17) & \multicolumn{1}{l}{Cls Ovr} & 1     & \multicolumn{1}{l}{S(-0.2)} & \multicolumn{1}{l}{Cor Rdn} & 1     & \multicolumn{1}{l}{N(-0.12)} &       &       &  \\
      & Cls Imb & 1     & N(-0.14) &       &       &       & \multicolumn{1}{l}{Mislabel} & 1     & \multicolumn{1}{l}{N(-0.11)} &       &       &  \\
      &       &       &       &       &       &       & \multicolumn{1}{l}{Cls Ovr} & 1     & \multicolumn{1}{l}{S(-0.16)} &       &       &  \\

\midrule

\multicolumn{1}{c}{\multirow{4}[0]{*}{\textbf{AU-ROC}}} & Tailed & -1    & S(0.17) & \multicolumn{1}{l}{Tailed} & -1    & \multicolumn{1}{l}{S(0.31)} & \multicolumn{1}{l}{Cor Rdn} & -1    & \multicolumn{1}{l}{S(0.27)} & \multicolumn{1}{l}{Mislabel} & -1    & \multicolumn{1}{l}{N(0.12)} \\
      & Cor Rdn & 1     & N(-0.12) & \multicolumn{1}{l}{Cls Ovr} & 1     & \multicolumn{1}{l}{\textbf{M(-0.47)}} & \multicolumn{1}{l}{Cls Ovr} & 1     & \multicolumn{1}{l}{\textbf{M(-0.44)}} & \multicolumn{1}{l}{Cls Ovr} & 1     & \multicolumn{1}{l}{\textbf{M(-0.36)}} \\
      & Cls Ovr & 2     & S(-0.3) &       &       &       &       &       &       &       &       &  \\
      & Cls Imb & -1    & N(0.13) &       &       &       &       &       &       &       &       &  \\
\bottomrule
\end{tabular}
\end{table}
\noindent \textbf{\textit{Results.}}
\textbf{Five antipatterns \textit{(Tailed, Mislabel, Class Overlap, Class Imbalance, correlation \& redundancy)} have a statistically significant effect on the performance metrics in the absence of other antipatterns; however, most antipatterns introduce only a small to medium boost/reduction of performance of ML models as indicated by the small to medium effect size.}
Table~\ref{tab.ap}\footnote{``N'', ``S'', ``M'' and ``L'' indicate negligible, small, medium, and large effect sizes respectively. The numerical value is provided next to the interpretation.}
\footnote{Antipatterns having medium and large effect sizes are bolded.}
\footnote{Negative sign shows increase in model performance after the introduction of that antipattern in the clean set.} shows the antipatterns\footnote{Abbreviations used: ``Cls" for ``Class", ``Ovr'' for ``Overlap'', ``Imb'' for ``Imbalance'', ``Cor Rdn'' for Correlation \& Redundancy, and ``Unnrm'' for ``Unnormalized''} that have a statistically significant comparison to the clean set accompanied by the cliff's effect size.

\textbf{Different antipatterns have different impacts on different performance metrics.} However, only the \textit{Tailed} and \textit{Class Overlap} antipatterns have a medium or large effect size for some performance metrics. \textbf{The Class Overlap antipattern has a statistically significant impact on multiple metrics with a small to large effect size.} Along with the AU-ROC metric, which has a medium effect size for three out of the four studied learners, class overlap impacts recall with medium to large effect size for two learners (NN and RF), medium effect on MCC for two learners (NN and LR), along with a small effect on the precision of RF and XGB. A negative effect size indicates that introducing the specific antipattern increase the model performance. Particularly, all four learners observe a boost in performance with class overlapped rows introduced into the dataset.
Similarly, \textbf{the introduction of Tailed distributions impacts the model performance metrics with a small to medium effect size}; however, this impact is metric-dependent as shown in the Table~\ref{tab.ap}. Other antipatterns like unnormalized distributions\footnote{While NN has an impact on the recall metric, the effect size is negligible}, schema violations, and duplicates do not have any statistically significant difference, and hence are absent from Table~\ref{tab.ap}. Interestingly, the mislabel antipattern had no effect on any performance metric except for Recall for NN and LR. We check the reasons for such a result in Section~\ref{sec.discussion}.

On the learner level, only the NN learner is impacted by class imbalance with a small effect size. The XGB learner is impacted by the least antipatterns (only Class Overlap and correlation \& redundancy), while other antipatterns have a negligible effect. Moreover, only the LR learner is impacted by Unnormalized features, however, the effect size is negligible. \\
Furthermore, on the learner level most learners assign the first SK rank to the class overlap antipattern for different metrics, indicating that all learners are in agreement that class overlap rows impact all learners' performance the most. Particularly, the learner's aggregated agreement on SKranks along different metrics for Class Overlap is 0.75; tailed is 0.79; constant feature, the clean set,  schema violations, and package antipatterns is 0.62; class imbalance and un-normal is 0.58; correlation \& redundancy is 0.54; and mislabel is the least of 0.5. Along with the results presented in Table~\ref{tab.ap}, this indicates that different antipatterns have different impact on the model performance. 

On a metric level, only the MCC for NN and XGB is impacted by Correlations \& Redundancies. Similarly, Mislabel impacts only the recall metric (while other effects are negligible). Finally, F1 is impacted by the least antipattern\footnote{Most antipatterns have negligible effect size.}.

\begin{Summary}{Summary of RQ3}{}
Although five antipatterns (\textit{Tailed, Mislabel, Class Overlap, Class Imbalance, Correlation \& redundancy}) have a statistically significant effect on performance metrics, most antipatterns only introduce a negligible to small effect. Among them, only the Tailed and Class Overlap antipatterns result in a medium or large effect on some of the model performance metrics.
\end{Summary}

%% file: rq3.tex
\subsection*{\rqthree}

\begin{figure}[!h]
      \makebox[\textwidth][c]{
      \includegraphics[width=1.2\columnwidth, keepaspectratio]{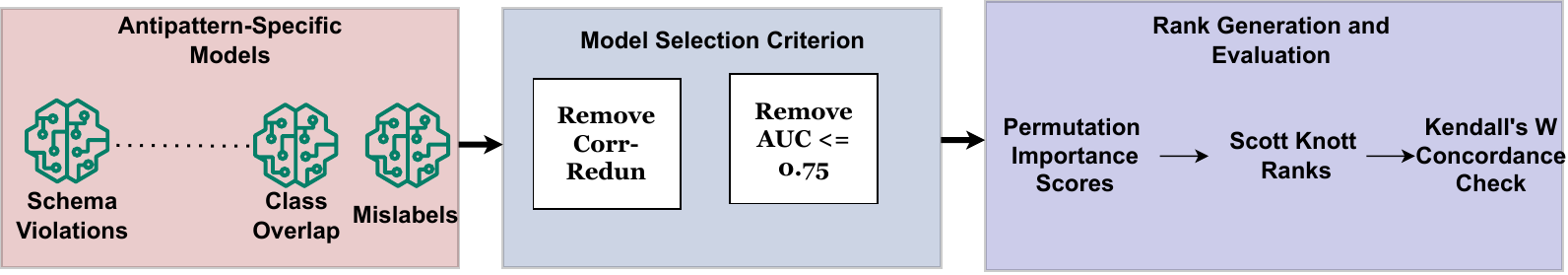}
      }
      \caption{Methodology to check concordance between different antipattern-specific models.}
      \label{fig.interpretation.methodology}
    \end{figure}
    
\textbf{\textit{Motivation.}} Interpretation of ML models is essential in various organizational tasks, such as defect prediction, to avoid misguided decisions. While interpretation remains a significant area of research, current research primarily focuses on the model perspective, including the impact of model performance~\cite{rajbahadur2021impact}, and the choice of learners~\cite{lyu2021empirical}. However, the influence of data quality on interpretation consistency has not been explored.

This research question aims to investigate the impact of antipatterns on interpretation consistency, thereby improving our understanding of the relationship between data quality antipatterns and the reliability of interpretation outcomes obtained from ML models.

\textbf{\textit{Approach.}} 
To check whether interpretation ranks from different antipatterns are consistent, we select the ML models from different antipatterns built in \textbf{RQ3}, and use permutation importance scores to calculate the feature importance obtained for each feature from the model. Finally, we measure the concordance between the interpretation ranks of models built from different antipatterns. We illustrate an overview of this methodology in Figure~\ref{fig.interpretation.methodology}, and provide the details below:
\begin{itemize}
    \item \textbf{Selection of ML models.} We use the ML models created for \textbf{RQ3}, each containing one antipattern at a time. Only models with AU-ROC $\geq 0.75$ are considered, as models with lower AU-ROCs produce inconsistent interpretation results~\cite{lyu2021towards,rajbahadur2021impact}. The threshold of 0.75 is also used by Lyu et al.~\cite{lyu2021towards} for predicting failures in job and hard drives. Moreover, we exclude the Correlation \& Redundancy antipattern from our analysis, as interpretation results from models with correlated \& redundant metrics are unreliable~\cite{rajbahadur2021impact}.
    \item \textbf{Generating Interpretation Scores.} From the models selected in the previous step, we extract the important features learned by that model using permutation importance~\cite{breiman2001random}. This study adopts a model-level interpretation approach, as opposed to instance-level interpretation. 
    
    \item \textbf{Generating Feature Importance Ranks.}
    The model interpretation is computed on a per-project-version-level and per-learner to account for inconsistencies in interpretation across different project versions and learners~\cite{lyu2021towards}. In particular, for each learner, we rank the permutation importance scores from models of each project-version to compute a feature importance rank. We use the Scott Knott ESD ranking algorithm for this purpose.

    \item \textbf{Evaluation of concordance.} We use Kendall's $W$ to calculate concordance between the interpretation ranks from different antipatterns. Particularly, we calculate the consistency of interpretation ranks from different antipatterns of the same project-version. Next, we aggregate the calculated Kendall's $W$ values for each of the four learners in a boxplot, and inspect the median values. 
    
    Kendall's $W$ is a non-parametric measure of inter-rater agreement and consistency, particularly in cases where there are multiple raters (i.e., ML models built from the eight antipatterns along with the clean set in our case). It takes into account the agreement among all raters, rather than just comparing pairs of raters or measures, making it a more comprehensive measure of agreement. This metric was also adopted by Lyu et al.~\cite{lyu2021towards} to determine the consistency of interpretation rankings from different learners. Kendall's $W$ ranges from -1 to 1, with values close to 1 indicating a strong positive association, values close to -1 indicating a strong negative association, and values close to 0 indicating no association. The Kendall's $W$ interpretation is as follows:
    \setlength{\abovedisplayskip}{0pt}
    \begin{align*}
    \text{Kendall's W Concordance} = 
        \begin{cases}
            \text{Weak,} & \text{if } | \text{W}|  \leq 0.3 \\
            \text{Moderate,} & \text{if }  0.3 <| \text{W} |\leq 0.6 \\
            \text{Strong} & \text{if } |\text{W}| > 0.6
        \end{cases}
    \end{align*}

    \item \textbf{Evaluation of pair-wise consistency} 
    Along with the internal consistency in-between the entire set of antipatterns, we then wish to check the consistency of interpretation results form an antipattern w.r.t. a standard set. We use the ``clean'' interpretation results for the data set to set as a baseline. While the correctness of the ``clean" dataset can be argued, our primary focus lies in the consistency of the interpretations rather than their absolute correctness. Hence, our selected ``clean" dataset acts merely as a reference, irrespective of potential discrepancies in the interpretation outcomes derived from this set.

    To serve our purpose of evaluation of pair-wise concordance, we use the Kendall's $\tau$ (unlike Kendall's $W$ which is used for multiple comparisons) metric. Kendal's $\tau$ is a non-parametric measure of similarity (consistency) between two rankings. Analogous to Kendall's $w$, it ranges between 0 (no agreement) to 1 (complete agreement). According to its interpretation, a weak agreement is indicated by values ranging from 0 to 0.3, a moderate agreement is denoted by values from 0.3 to 0.6, and a strong agreement is represented by values from 0.6 to 1.

    Similar to Kendall's $W$, the $\tau$ value is calculated on a per-project level basis from the ML models built from \textbf{RQ3}.
    
\end{itemize}

\noindent \textbf{\textit{Results.}}

    \begin{figure}[!t]
      \centering
      \includegraphics[width=0.9\columnwidth, keepaspectratio]{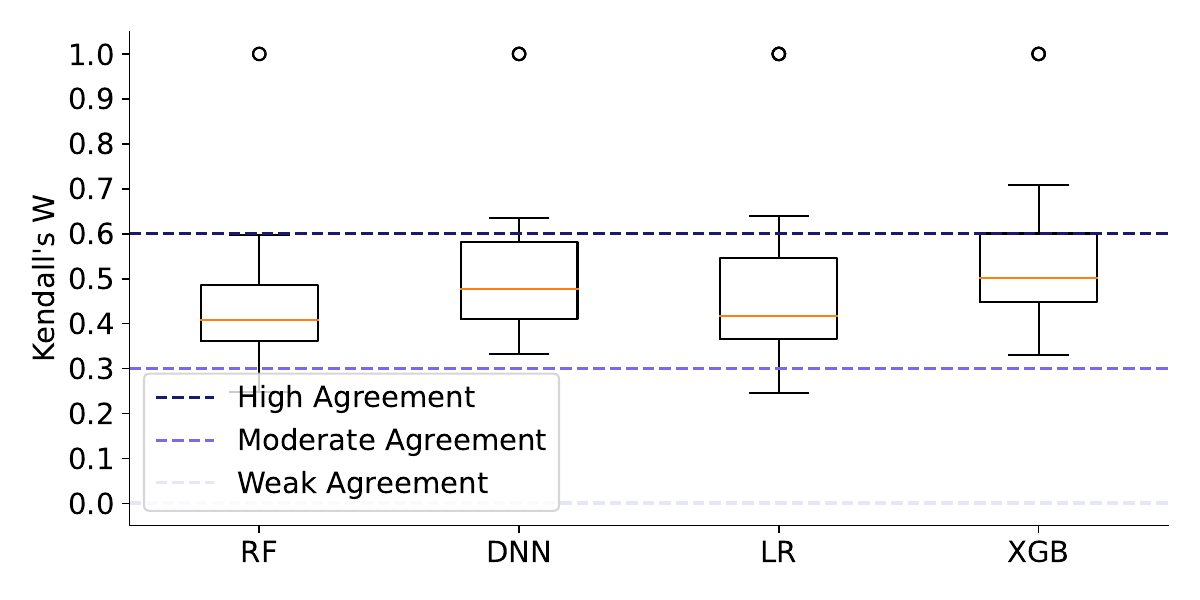}
      \caption{Kendal's $W$ concordance for Interpretation Ranks from multiple antipatterns.}
      \label{fig.interpretation.KendalW}
    \end{figure}

    \textbf{Interpretations from models with different antipatterns show moderate consistency.} Figure~\ref{fig.interpretation.KendalW} displays the concordance of interpretation ranks among various antipattern-specific models for each project version. For all learners, the median concordance falls between the 0.6 and 0.3 marks, as depicted in the figure. This result suggests that there is a moderate level of agreement in the interpretation ranks across models with different antipatterns. 

\begin{figure}[!h]
      \centering
      \makebox[\textwidth][c]{\includegraphics[width=1.3\columnwidth, keepaspectratio]{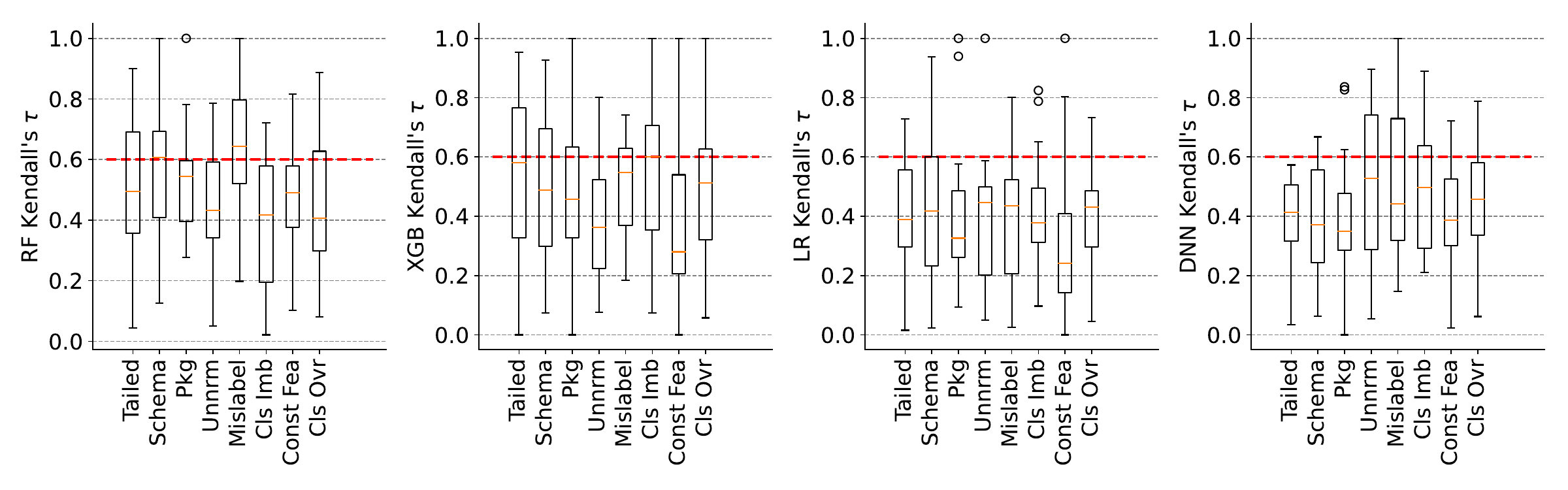}}
      \caption{Paired correlation of interpretation ranks from models trained with different antipatterns w.r.t the clean set.}
      \label{fig.interpretation.KendalTau}
    \end{figure}

    Furthermore, our results of pair-wise concordance checks are presented in Figure~\ref{fig.interpretation.KendalTau}. We observe that pair-wise consistency of interpretation results from different antipatterns with respect to the clean set is in-between 0.3 to 0.6, which is again moderate. The moderate consistency obtained from Figure~\ref{fig.interpretation.KendalTau} further strengthens the moderate consistency in-between antipatterns findings from Fig.~\ref{fig.interpretation.KendalW}.

\begin{Summary}{Summary of RQ4}{}
Data quality antipatterns render the interpretation results moderately inconsistent. While prior research has uncovered factors like learner characteristics, project specifics, hyperparameter tuning, and sampling randomness as key influencers of interpretation consistency, our research is the first to investigate the impact of ``data quality'' on the interpretation of machine learning models. We provide empirical evidence showing that data quality contributes to moderate inconsistencies in interpretation results.
\end{Summary}

%% file: Threats.tex
\section{Threats to Validity}\label{sec.threats}

\subsection{Threats to External Validity}

Our selected SDP dataset with 30 project versions provided us with substantial volume of data to train models for four learners. Overall, the sheer size of our dataset and complexity of our experiments led us to train 26,400 ML trained models, which is non-trivial. Despite the extensive experimentation within a single domain (SDP), like all research, our results are inherently limited by the specific datasets and the four learners used. Furthermore, to minimize this threat for learners, we selected learners commonly used in defect prediction research and represent a diverse range of families~\cite{tantithamthavorn2018experience}. Other researchers may need to apply our taxonomy to their datasets and contexts expand upon the generalizability of our findings and verify the impact of antipatterns in different environments.

Moreover, in RQ2 and RQ3, we used Sklearn version 1.1.3 to build our models. This framework is popular and widely adopted, ensuring consistency across our experiments. However, the exclusion of other frameworks such as R, C++, or Java could limit the broader applicability of our findings. Future studies might explore different tools and frameworks to enhance the generalizability of the results. \\

\subsection{Threats to Internal Validity} In \textbf{RQ3}, we examine the effect of a specific antipattern by introducing it to the clean dataset. This experiment design allows us to evaluate the impact of the antipattern in an isolated setting. An alternative experiment design where all antipatterns are added and then one is removed could have been conducted. However, the outcome of such an experiment would reveal the impact of ``cleaning'' the antipattern, rather than the antipattern itself.\\
Moreover, in \textbf{RQ4}, the choice of AU-ROC threshold (0.75) may impact our results. Yet, a high AU-ROC threshold is necessary to evaluate the impact inconsistency caused by low-performing models~\cite{lyu2021towards,rajbahadur2021impact}. Furthermore, detecting data drift involves experimenting with various thresholds, and while we found no significant drift between subsequent versions, this process could be influenced by threshold selection. We adopted the default values provided by Caveness et al.~\cite{caveness2020tensorflow} to detect drift in our dataset. Along similar lines, the high variance for the number of correlated \& redundant features might have been introduced by Jiarpakdee et al.`s tool~\cite{jiarpakdee2018autospearman}, which may have selected different features from the correlated and redundant metrics. Although the Jiarpakdee et al.`s research focused finding consistent metrics for SDP, future research should study the consistency of metrics across different project-versions. 


\subsection{Threats to Construct Validity}
For \textbf{RQ2} and \textbf{RQ3}, we conduct the model training process 10 times for each learner for each order/antipattern. While other studies have utilized a higher number of repetitions to broaden the number of randomly sampled data splits, the sheer number of ML models generated in this research is substantial. In particular, a total of $12(orders)*4(learners)*30(project-versions)*10(runs)=14,400$ models were generated for \textbf{RQ2} and $10(antipatterns)*4(learners)*30(project-versions)*10(runs)=12,000$ models were generated for \textbf{RQ3}, resulting in a grand total of 26,400 models. All models are generated using hyperparameter tuning with 10 iterations and cross validation parameter   3, amounting to training a total of 0.78 Million models. Despite the use of high-performance computing resources (on a server machine with 80 CPU cores), the sheer volume of ML models generated resulted in a substantial computing time ranging ~7 days.


Additionally, while prior research~\cite{bergstra2012random} has indicated that random search is superior to other optimization techniques such as grid search, we cannot guarantee that our set of extracted hyperparameters are the most optimal.

To gauge the consistency of interpretation in \textbf{RQ4}, the use of permutation importance as the sole feature importance measure may not capture the full complexity of model interpretation. Future research may wish to employ alternative feature importance measures, such as LIME \cite{ribeiro2016should} or SHAP, to provide a more comprehensive understanding of the consistency of interpretation results from different antipatterns.
Moreover, in \textbf{RQ4}, we utilize the Kendall's $W$ metric. While research~\cite{rajbahadur2021impact,lyu2021empirical} dealing with low features have used metrics like Top-N score, we believe Kendall's W metric provides the most appropriate concordance results due to a high feature count in our SDP dataset. We believe that the Top-N metric would be beneficial for lower dimensional datasets as used by Lyu et al.~\cite{lyu2021empirical}.

%% file: Discussion.tex
\section{Discussion}\label{sec.discussion}

In this section, we provide a more detailed analysis of the results obtained in \textbf{RQ3}. Specifically, we did not observe any impact of mislabels on the performance of SDP models. Since this result contradicts prior research~\cite{yatish2019mining}, we conduct further analysis to support our findings. Thus, in this subsection, we examine the inter-relationships between different antipatterns to explain our \textbf{RQ3} finding that Mislabeling has no effect on model performance.

\noindent\textbf{Approach.} In \textbf{RQ1}, we found that 94\% of row-level antipattern overlaps are due to \textit{Mislabel} and \textit{Class Overlap} antipatterns. This indicates a need to examine the impact of class overlap on mislabel correction. Hence, we obtained AU-ROC results from models trained on five scenarios: \textit{FiTrOv}, where mislabel correction is not performed; \textit{FiTrMi}, where mislabel correction is performed but class overlapping rows are not removed; \textit{FiTrOvMi}, where both class overlapping rows and mislabels are removed; \textit{FiTrMiOv}, where both class overlapping rows and mislabels are removed, but the removal of class overlapping rows is based on incorrect labels; and \textit{FiTr}, where neither class overlapping rows nor mislabels are corrected. 
Since Filter and Transformation based antipatterns do not have much significant impact from our \textbf{RQ3} findings, our \textit{FiTr} and \textit{FiTrMi} cases are comparable to Yatish et al.'s~\cite{yatish2019mining} experimental setup. To compare our findings to theirs, we only used the AU-ROC metric in this subsection.

\begin{figure}[!h]
\centering
      \makebox[\textwidth][c]{\includegraphics[width=1.2\columnwidth, keepaspectratio]{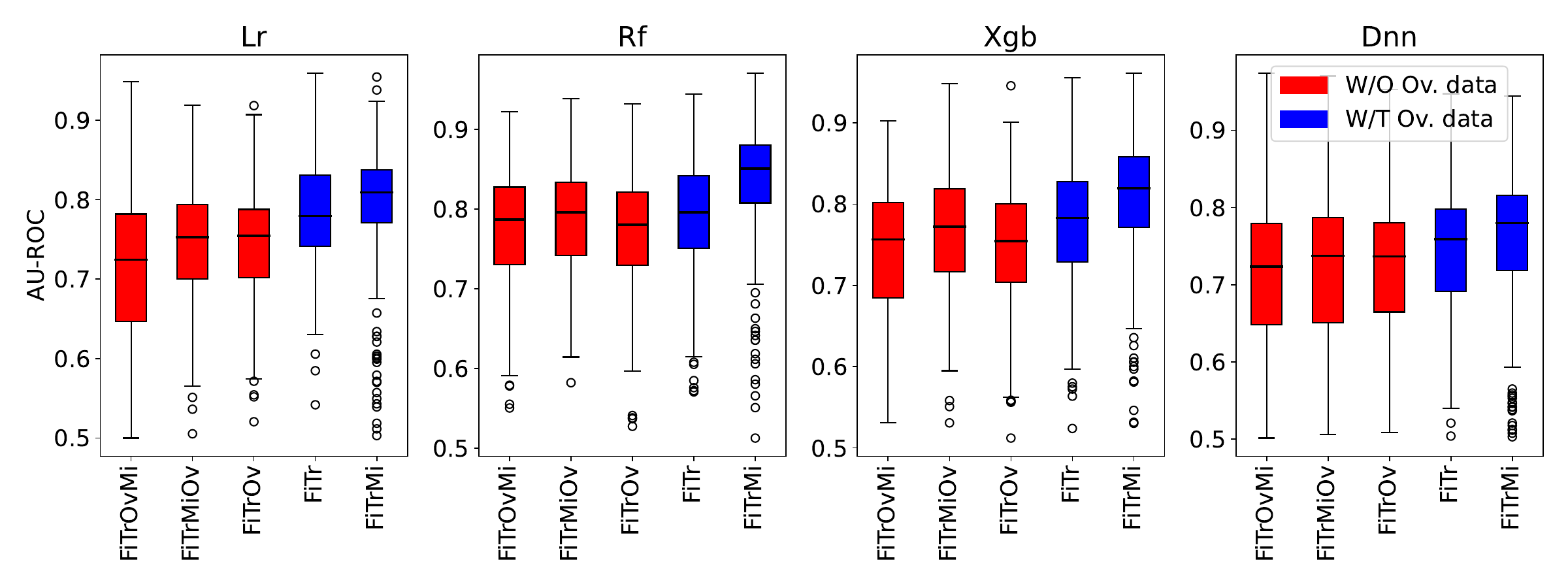}}
\caption{Impact of Class Overlap removal on Mislabels. For ease of interpretation, red boxes show the removal of class overlapped rows, while the two blue boxes show the replication of Yatish's~\cite{yatish2019mining} experiments with (`W/T' represented by blue) and without (`W/O' represented by red) Mislabels.}
\label{fig.mislabeloverlap}
\end{figure}

\textbf{Results: Removing class-overlapping rows overshadows the impact of mislabeling.} Figure~\ref{fig.mislabeloverlap} shows orders with class-overlapping rows removed in red and orders with class-overlapping rows present in blue. The blue boxes indicate that mislabeling significantly affects the model's AU-ROC, as previously identified by Yatish et al.~\cite{yatish2019mining}. The order \textit{FiTr} exhibits a lower AU-ROC than the order \textit{FiTrMi}, where mislabel correction is performed. This observation is supported by statistically significant comparisons (Wilcoxon Rank-Sum test, $p-value \leq 0.05$) with small to medium effect sizes.

However, when class-overlapping rows are removed (i.e., red boxes in Figure~\ref{fig.mislabeloverlap}), the orders \textit{FiTrOvMi} and \textit{FiTrMiOv} do not exhibit a very distinct AU-ROC compared to \textit{FiMiOv}, which demonstrates the diminished impact of mislabeling. This is further evidenced by statistically insignificant differences (Wilcoxon Rank-Sum test, $p-value \leq 0.05$) between \textit{FiTrOv} and \textit{FiTrOvMi} with a medium to large effect sizes for three out of four classifiers, accompanied by a small effect size (Cliff's $\delta = -0.19$) for the remaining classifier, Logistic Regression.

%% file: Implications.tex
\section{Implications}\label{sec.implications}

We present the key contributions of our study below: 

\begin{itemize}
    \item Our \textbf{taxonomy} of data quality antipatterns, which includes eight main types and fourteen sub-types, serves as a critical resource for researchers and practitioners aiming to evaluate and mitigate antipatterns in their datasets. This comprehensive checklist aids in identifying prevalent data quality issues, which can significantly impact the performance of machine learning (ML) models. The taxonomy facilitates a systematic approach to data cleaning, enabling data scientists to prioritize and address specific antipatterns relevant to their use case, thereby enhancing the reliability and effectiveness of their ML models.

    \item Our finding that the order of addressing antipatterns can significantly affect model performance is important for practitioners. Specifically, handling class-based antipatterns like class overlap should be done towards the end, especially in large-scale pipelines with data from diverse sources where mislabels can be prevalent. In scenarios where adjusting the cleaning order is not feasible, opting for robust learners such as deep neural networks (DNNs) is recommended due to their resilience to different cleaning orders as indicated by our results. Practitioners can take inspiration from our results and minimize performance degradation caused by inaccurate order of cleaning antipatterns while building their ML models.

    \item Our research evaluates the impact of data quality antipatterns on model performance in real-world settings by assessing the effect of adding antipatterns to a clean dataset to isolate their effects. This approach differs from other studies that perform ablation experiments by removing antipatterns one at a time. Ablation experiments may not fully reveal the impact of individual antipatterns since multiple antipatterns often coexist, leading to interlinked effects. For example, removing class overlap could inadvertently address class imbalance issues. While we highlight this crucial aspect, future research should further analyze the interdependence of antipatterns.

    Our research is particularly relevant to real-world settings where data procurement is distributed, such as in federated learning, where cleaning occurs at multiple phases, devices, and times. Given that some antipatterns are already addressed in one phase, removing additional antipatterns in a partially cleaned dataset may not be beneficial since antipatterns have overlapping effects, and preserving the data characteristics may be more important than culling additional ``perceived'' noise from a dataset that is already partially cleaned out.
    
    Moreover, our results are also relevant to scenarios where limited data exists for model training. Since the performance of ML models heavily depends on data size, sufficiently cleaned data may be adequate. Overzealous cleaning can remove valuable signals, mistaking them for noise. Specifically, our findings indicate that in partially-cleaner settings, class overlap may remove essential features needed for model decision-making, emphasizing the importance of strategic cleaning decisions.

    \item Our \textbf{schema} for the SDP dataset assists future SDP researchers and practitioners in addressing schema violations in their datasets. By developing a schema for the SDP dataset, we provide a framework that not only aids software researchers but also encourages data scientists and engineers to identify and rectify questionable or incorrect values in their datasets. Future research and practitioners can draw inspiration from our study to create a schema for their datasets, thereby improving data quality and model performance.
    
\end{itemize}

%% file: Conclusion.tex
\section{Conclusion}\label{sec.conclusion}

The quality of data utilized in training machine learning pipelines is a burgeoning area of research. While previous studies have examined individual data quality antipatterns, it is plausible that multiple antipatterns coexist within a dataset, potentially compounding their effects and that they present complexities like unknown order of their removal. To address real-world  data cleaning challenges during ML pipeline training, we first introduce a comprehensive taxonomy of data quality antipatterns for structured data, comprising of nine top-level and 14 sub-types. This taxonomy serves as a foundational framework for identifying and mitigating data quality issues across various domains.

In this research, we evaluate the impact of these antipatterns by specifically focusing on the domain of Software Defect Prediction (SDP), a crucial area within software engineering that forecasts potential defects in software applications, thereby maintaining software quality and reducing maintenance costs. Highlighting SDP underscores its significance as the canonical software analytics application of ML within software engineering and provides a concrete case study for the broader implications of data quality antipatterns.

Our findings indicate that nine out of the 19 identified antipatterns (five top-level and 14 sub-types) are prevalent in our studied SDP dataset. Five antipatterns—\textit{tailed distributions, mislabels, class overlap, class imbalance}, and \textit{correlation \& redundancy}—significantly impact model performance. Overall there is substantial variability in the impact of different antipatterns across different learners, indicating that different antipatterns impact different learners differently. Software researchers and practitioners must be vigilant about the order of cleaning different antipatterns, as different orders of cleaning can statistically significantly impact model performance. This can be particularly critical when dealing with distributed or large-scale data pipelines, where different cleaning steps may occur at various stages or locations of data collection and curation. Furthermore, our study is the first to explore the impact of data quality on model interpretation, revealing that these antipatterns can moderately affect interpretive consistency. Particularly, we highlight the importance of considering data quality while entrusting the interpretation results of software analytics models.

Overall, our research contributes to the literature on data quality by addressing data engineering challenges in software defect prediction and offering novel insights into how data quality antipatterns affect model performance and interpretation. Software researchers should replicate our study to other software analytics tasks, while practitioners should enrich the development of their data cleaning tools. Furthermore, researchers and practitioners from various domains can leverage our antipattern taxonomy towards a systematic approach for cleaning their datasets.